\documentclass[12pt]{article}
\usepackage{epsfig,amssymb,amsmath,psfrag}
\usepackage{tikz}
\DeclareMathOperator{\Li}{Li}
\usetikzlibrary{decorations.pathmorphing} 

\def \bl  {\begin{align*}}
\def \el  {\end{align*}}

\def \be  {\begin{equation}}
\def \ee  {\end{equation}}
\def \ba  {\begin{eqnarray}}
\def \ea  {\end{eqnarray}}
\def \baa {\begin{eqnarray*}}
\def \eaa {\end{eqnarray*}}
\def \bb  {\begin {thebibliography} }
\def \eb  {\end{thebibliography}}
\def \lab #1 {\label{#1}}

\newcommand{\beq}{\begin{equation}}
\newcommand{\eeq}{\end{equation}}
\newcommand{\beqa}{\begin{eqnarray}}
\newcommand{\eeqa}{\end{eqnarray}}

\def \tr {\mathop{\rm tr}\nolimits}

\def\l<{\langle}
\def\r>{\rangle}

\def\XXint#1#2#3{{\setbox0=\hbox{$#1{#2#3}{\int}$}
     \vcenter{\hbox{$#2#3$}}\kern-.5\wd0}}

\renewcommand{\title}[1]{\vbox{\center\LARGE{#1}}\vspace{5mm}}
\renewcommand{\author}[1]{\vbox{\center#1}\vspace{5mm}}

\usepackage{pgfkeys}
\usepackage{pgffor}
\usepackage{tikz}
\usepackage{ifthen}
\usetikzlibrary{intersections}
\usetikzlibrary{decorations.pathmorphing}
\usetikzlibrary{calc}

\tikzset{%
  set style/.style={#1},
  expand style/.style={
    set style/.expanded={#1}
  },
  expand style once/.style={
    set style/.expand once={#1}
  },
  expand style twice/.style={
    set style/.expand twice={#1}
  }
}

\pgfkeys{
  /ww/.is family,
  /ww/.unknown/.code = {} 
}

\ExplSyntaxOn
\cs_new_eq:NN \ipLoop \int_step_inline:nn
\ExplSyntaxOff

\pgfkeys{
 /ww/defineline/.is family,/ww/defineline/.cd,
  /ww/defineline/.unknown/.code = {},/ww/defineline/linelength/.store in =\linelength, /ww/defineline/linelength/.default =1%
}

\newcommand{\defineline}[2]{%

  \pgfkeys{/ww/defineline/.cd,linelength}
  \pgfkeys{/ww/defineline/.cd,#2} 

  \pgfkeys{/ww/#1/.is family}%
  \pgfkeys{/ww/#1/.unknown/.code = \pgfkeyssetvalue{/ww/#1/\pgfkeyscurrentname}{##1}}%

  \pgfkeys{/ww/#1/.cd, #2}%

  \pgfkeysgetvalue{/ww/#1/pointcount}{\pointcount}%
  \def\pointsrange{(1.4*\linelength)}

  \pgfmathparse{\pointsrange/(\pointcount-1)}
  \def\pointgap{\pgfmathresult}
  
  \ipLoop{\pointcount}{

    \pgfkeysgetvalue{/ww/#1/originx}{\originx}
    \pgfkeysgetvalue{/ww/#1/originy}{\originy}
    \pgfkeysgetvalue{/ww/#1/angle}{\angle}

   \pgfmathsetmacro{\pointx}{\originx +(-\pointsrange/2 + (##1-1)*\pointgap)*cos(\angle)}   
   \pgfmathsetmacro{\pointy}{\originy +(-\pointsrange/2 + (##1-1)*\pointgap)*sin(\angle)}
   \pgfmathsetmacro{\linelengthtwo}{\linelength}

   \edef\keypx{/ww/#1/ipx##1}
   \edef\keypy{/ww/#1/ipy##1}
   \pgfkeyslet{\keypx}{\pointx}
   \pgfkeyslet{\keypy}{\pointy}

   \edef\keylinelength{/ww/#1/ilinelength}
   \pgfkeyslet{\keylinelength}{\linelengthtwo}

  }

}

\tikzset{gluon/.style={decorate, decoration={coil,amplitude=2pt, segment length=4pt},color=purple}}

\pgfkeys{
  /ww/prop/.cd,
  /ww/prop/.is family,
  /ww/prop/fromline/.initial = 1,
  /ww/prop/fromline/.store in = \fromline,
  /ww/prop/toline/.initial = 1,
  /ww/prop/toline/.store in=\toline,
  /ww/prop/fromip/.initial = 1,
  /ww/prop/fromip/.store in=\fromip,
  /ww/prop/toip/.initial = 1,
  /ww/prop/toip/.store in=\toip,
  /ww/prop/stublength/.store in=\stublength,
  /ww/prop/stublength/.default = 0.5,
  /ww/prop/markendpoints/.store in = \markendpoints,
  /ww/prop/markendpoints/.default = false,
  /ww/prop/startpointlabel/.store in=\startpointlabel,
  /ww/prop/startpointlabel/.default = \phantom{.},  
  /ww/prop/endpointlabel/.store in=\endpointlabel,
  /ww/prop/endpointlabel/.default = \phantom{.},  /ww/prop/labeldistone/.store in = \labeldistone,
  /ww/prop/labeldistone/.default = 0.5,
  /ww/prop/labeldisttwo/.store in = \labeldisttwo,
  /ww/prop/labeldisttwo/.default = 0.5,
  /ww/prop/looseval/.store in = \looseval,
  /ww/prop/looseval/.default = 1.4
}%

\pgfkeys{
  /ww/drawline/.is family,/ww/drawline/.cd,
  /ww/drawline/.unknown/.code = {},%
  /ww/drawline/customstyle/.initial = {line width=1},
  /ww/drawline/customstyle/.default = {line width=1},
  /ww/drawline/showtwistors/.store in = \showtwistors,
  /ww/drawline/showtwistors/.default = false,
  /ww/drawline/showtwistorsout/.store in = \showtwistorsout,
  /ww/drawline/showtwistorsout/.default = false,
  /ww/drawline/showwings/.store in = \showwings,
  /ww/drawline/showwings/.default = false,
  /ww/drawline/showvertex/.store in = \showvertex,
  /ww/drawline/showvertex/.default = false,
  /ww/drawline/labeldist/.store in = \labeldist,
  /ww/drawline/labeldist/.default = 0.5,
  /ww/drawline/wingorientation/.store in = \wingorientation,
  /ww/drawline/wingorientation/.default = inner,
  /ww/drawline/tolabel/.store in = \tolabel,
  /ww/drawline/tolabel/.default = ,
  /ww/drawline/fromlabel/.store in = \fromlabel,
  /ww/drawline/fromlabel/.default = ,
  /ww/drawline/linelength/.store in = \linelength,
}

\newcommand{\drawline}[2][]{%
  \pgfkeys{/ww/drawline/.cd,tolabel,fromlabel,customstyle,showtwistors,showtwistorsout,showwings,wingorientation,showvertex,labeldist}
  \pgfkeys{/ww/drawline/.cd,#1}
  
  \pgfkeysgetvalue{/ww/#2/originx}{\ox}%
  \pgfkeysgetvalue{/ww/#2/originy}{\oy}%
  \pgfkeysgetvalue{/ww/#2/angle}{\ang}%
  \pgfkeysgetvalue{/ww/#2/ilinelength}{\linelength}%

  \pgfmathsetmacro{\oxf}{\ox}%
  \pgfmathsetmacro{\oyf}{\oy}%
  \pgfmathsetmacro{\angf}{\ang}%
  \pgfmathsetmacro{\linelengthf}{\linelength}

  \pgfkeysgetvalue{/ww/drawline/customstyle}{\customstyle}
  
  \draw[expand style = \customstyle, name path=#2]
    ($(\oxf,\oyf) - (\angf:\linelengthf+0.3)$) 
    -- 
    ($(\oxf,\oyf) + (\angf:\linelengthf+0.3)$);

    \ifthenelse{\equal{\showtwistors}{true}}
           {
           
           \ifthenelse{\equal{\showtwistorsout}{false}}{\path ($(\oxf,\oyf) + (\angf:\linelengthf+0.0) + (\angf+90:\labeldist)$) node {\fromlabel};
    \path ($(\oxf,\oyf) - (\angf:\linelengthf+0.0) + (\angf+90:\labeldist)$) node {\tolabel}}{\path ($(\oxf,\oyf) + (\angf:\linelengthf+0.3) + (\angf+90:\labeldist)$) node {\fromlabel};
    \path ($(\oxf,\oyf) - (\angf:\linelengthf+0.3) + (\angf+90:\labeldist)$) node {\tolabel}};

    \ifthenelse{\equal{\showvertex}{true}}{
    \draw[fill=black] ($(\oxf,\oyf) - (\angf:\linelengthf)$) circle (0.05);
    \draw[fill=black] ($(\oxf,\oyf) + (\angf:\linelengthf)$) circle (0.05);
    }
    }{}

    \ifthenelse{\equal{\showwings}{true}}
    {
      \ifthenelse{\equal{\wingorientation}{inner}}
      {
        \draw[expand style = \customstyle] ($(\oxf,\oyf) - (\angf:\linelengthf) - (\angf-30:-0.2) $) -- ($(\oxf,\oyf) - (\angf:\linelengthf) - (\angf-30:0.5)$);
        \draw[expand style = \customstyle] ($(\oxf,\oyf) + (\angf:\linelengthf) - (\angf-150:-0.2)$) -- ($(\oxf,\oyf) + (\angf:\linelengthf) - (\angf-150:0.5)$);
      }{
        \draw[expand style = \customstyle] ($(\oxf,\oyf) - (\angf:\linelengthf) - (\angf+30:-0.2) $) -- ($(\oxf,\oyf) - (\angf:\linelengthf) - (\angf+30:0.5)$);
        \draw[expand style = \customstyle] ($(\oxf,\oyf) + (\angf:\linelengthf) - (\angf+150:-0.2)$) -- ($(\oxf,\oyf) + (\angf:\linelengthf) - (\angf+150:0.5)$);
        }
    }{}

}%

\newcommand{\drawpropagator}[1]{

  \pgfkeys{/ww/prop/.cd,markendpoints,startpointlabel,endpointlabel,labeldistone,labeldisttwo,looseval}
  \pgfkeys{/ww/prop/.cd,#1}

  \pgfkeysgetvalue{/ww/\fromline/ipx\fromip}{\ix}%
  \pgfkeysgetvalue{/ww/\fromline/ipy\fromip}{\iy}%
  \pgfkeysgetvalue{/ww/\toline/ipx\toip}{\fx}%
  \pgfkeysgetvalue{/ww/\toline/ipy\toip}{\fy}%

  \pgfkeysgetvalue{/ww/\fromline/angle}{\angi}%
  \pgfkeysgetvalue{/ww/\fromline/pointcount}{\pointcount}%
  \pgfkeysgetvalue{/ww/\toline/angle}{\angt}%

  \pgfmathsetmacro{\angout}{\angi - 90}
  \pgfmathsetmacro{\angin}{\angt - 90}

  \draw[/tikz/gluon] (\ix,\iy) to[out=\angout, in = \angin,looseness=\looseval] (\fx,\fy);

  \ifthenelse{\equal{\markendpoints}{true}}
  {
    \path ($(\fx,\fy) + (\angin:-\labeldistone) $) node {\endpointlabel};
    \path ($(\ix,\iy) + (\angout:-\labeldisttwo) $) node {\startpointlabel};
  }

}%

\newcommand{\drawpropagatorstub}[1]{
  \pgfkeys{/ww/prop/.cd,markendpoints,startpointlabel,labeldistone,stublength}
  \pgfkeys{/ww/prop/.cd,#1}
  
  \pgfkeysgetvalue{/ww/\fromline/ipx\fromip}{\ix}%
  \pgfkeysgetvalue{/ww/\fromline/ipy\fromip}{\iy}%

  \pgfkeysgetvalue{/ww/\fromline/angle}{\angi}%
  \pgfkeysgetvalue{/ww/\fromline/pointcount}{\pointcount}%

  \pgfmathsetmacro{\angout}{\angi - 90}

  \draw[/tikz/gluon] (\ix,\iy) -- ++ (\angout:\stublength);

  \ifthenelse{\equal{\markendpoints}{true}}
  {
    \path ($(\ix,\iy) + (\angout:-\labeldistone) $) node {\startpointlabel};
  }
  
}%

\numberwithin{equation}{section}
\begin{document}

\thispagestyle{empty}

\begin{flushright}

\end{flushright}

\vskip2.2truecm
\begin{center}
\vskip 0.2truecm {\Large\bf
{\Large Light-like Wilson loop correlators and the $\bar{Q}$-equation}}\\
\vskip 1truecm
\vskip 1truecm
{\bf J.M. Drummond\footnote{J.M.Drummond@soton.ac.uk}, M. Rochford\footnote{M.J.Rochford@soton.ac.uk}, R. Wright\footnote{Rowan.Wright@soton.ac.uk}
}
\vskip 0.4truecm
{\emph{School of Physics and Astronomy, University of Southampton, Southampton, SO17 1BJ, UK
}}

\begingroup\bf\large

\endgroup
\vspace{0mm}

\begingroup
\textit{
 }\\
\par
\endgroup

\end{center}

\vskip 1truecm 
\centerline{\bf Abstract} 
In recent work we began a study of the correlators of multiple light-like Wilson loops in $\mathcal{N}=4$ super Yang-Mills theory, focussing primarily on tree-level calculations and, beyond tree-level, to the Abelian theory. Here we calculate $O(g^2)$ correlators of multiple light-like Wilson loops in the $SU(N)$ theory. We use the \emph{chiral box expansion} and a study of the leading singularities of the loop integrand to arrive at integrated expressions for these objects. We then use the results of these calculations to verify that a natural generalisation of the $\bar{Q}$-equation, familiar from the study of single Wilson loops, holds in the $SU(N)$ theory. This $\bar{Q}$-equation should provide a valuable tool for the computation of multiple Wilson loop correlators at higher order in the coupling. 

 \noindent

\newpage
\tableofcontents
\newpage
\setcounter{page}{1}\setcounter{footnote}{0}

\section{Introduction}
The expectation values of light-like Wilson loops are well-known to be dual to maximally helicity-violating scattering amplitudes in planar \(\mathcal{N}=4\) super Yang-Mills theory \cite{Alday:2007hr,Drummond:2007aua,Brandhuber:2007yx,Drummond:2007cf,Bern:2008ap,Drummond:2008aq}. Given the significant triumphs in the study of planar \(\mathcal{N}=4\) scattering amplitudes, and the fact that much of the rich mathematical structure observed for amplitudes is at least as manifest on the Wilson loop side of the duality, it is natural to investigate the extent to which other objects related to Wilson loops exhibit the same mathematical properties and structures. One obvious candidate is the correlator of a Wilson loop with the chiral Lagrangian of the theory, which has been the focus of much recent study \cite{Alday:2011ga,Alday:2012hy,Alday:2013ip,Chicherin:2022bov,Chicherin:2022zxo,Chicherin:2024hes,Carrolo:2025pue}. Here, we focus on another natural generalisation: the correlation functions of \emph{multiple} light-like (super) Wilson loops. 

In \cite{Drummond:2025ulh}, we presented some simple calculations of these objects at tree-level for the $SU(N)$ theory. We primarily used the twistor super Wilson loop formalism of Mason and Skinner \cite{Mason:2010yk}, and we refer the reader to that work for a detailed review of this formalism which we will only briefly recap here. We also presented a natural generalisation of the \(\bar{Q}\)-equation for correlators of multiple loop operators. This equation was originally derived for planar \(\mathcal{N}=4\) scattering amplitudes via Wilson loop arguments \cite{Caron-Huot:2011dec,Bullimore:2011kg}. The generalised \(\bar{Q}\)-equation takes the form
\be
\label{Qbargeneralm}
\bar{Q}_A^{A'} \mathcal{R}^{(k)}_{n_1,\ldots,n_m} = \frac{1}{4}\Gamma_{\rm cusp}\sum_r \biggl[   \int \bigl[d^{2|3} \mathcal{Z}_{n_r+1} \bigr]^{A'}_{A} \Bigl[ \mathcal{R}^{(k+1)}_{n_1,\ldots,n_r+1,\ldots,n_m} - \mathcal{R}^{(k)}_{n_1,\ldots,n_m} \mathcal{R}^{(1,0)}_{n_r+1}\Bigr] + \text{cyc}_r\biggr]\,.
\ee
We demonstrated in \cite{Drummond:2025ulh} that this equation holds in the Abelian theory and also at large $N$ for the leading non-trivial connected contributions in the $SU(N)$ theory which factorise into pieces coming from the Abelian theory.\footnote{For $m$ loop operators such leading connected contributions in the N${}^k$MHV sector are $O(g^{2(m-k)})$ for $k \leq m$.} Here, we perform explicit computations at $O(g^2)$ in the $SU(N)$ theory and demonstrate that this conjectured equation correctly relates integrated \(O(g^2)\) correlators to tree level i.e. \(O(g^0)\) correlators. In doing so, we use the \emph{chiral box expansion} introduced in \cite{Bourjaily:2013mma} to perform the loop integration, and utilise compact expressions for the \(O(g^2)\) leading singularities (which provide the coefficients of the chiral box integrals) in terms of simple, tree-level objects. We will discuss the general form of the leading singularities of multiple light-like Wilson loop correlators at \(O(g^2)\) in more detail in a companion paper \cite{leadingSing}.

Although it is in principle straightforward to generate the loop integrand of a multiple Wilson loop correlator at any loop order, beyond \(O(g^2)\) the loop integration becomes challenging, as is the case for the expectation value of a single Wilson loop (see \cite{Spiering:2024sea} for results on two-loop integrals that appear in the planar $\mathcal{N}=4$ SYM amplitudes). In that case, the original \(\bar{Q}\)-equation has provided a vital tool which has allowed computations to be performed even at \(O(g^6)\) and \(O(g^4)\) in the case of an octagon and nonagon respectively \cite{Li:2021bwg,He:2020vob}. The existence of a \(\bar{Q}\)-equation for multiple Wilson loop correlators suggests that computations at high loop order may be tractable through the same procedure, although we defer pursuit of such calculations to future work. 

\newpage

\section{Setup and conventions}

\subsection{Multiple light-like Wilson loop correlators}
We perform our calculations in the context of four-dimensional, maximally supersymmetric Yang-Mills theory with action
\be
S = \frac{1}{g_{\rm YM}^2} \int d^4x \biggl[ -\frac{1}{2} \tr F^{\mu \nu} F_{\mu \nu}  + \ldots\biggr]\,.
\ee
where we have omitted the fermionic and scalar fields, and have normalised the fields such that the coupling \(g_{\rm YM}^2\) only appears as a pre-factor in front of the action. Here we will consider the gauge group $SU(N)$ although the computations presented generalise to other cases e.g. $U(N)$ and including the Abelian $U(1)$ theory. In performing calculations, we will make use of the t'Hooft coupling,
\be
g^2 = \frac{g^2_{\rm YM}N}{16\pi^2}.
\ee
The objects we study here are the expectation values and correlation functions of loop operators
\be
\mathcal{L}(C) = \frac{1}{N} \tr \mathcal{P}\,{\rm exp} \,i  \oint_C  dx^{\mu}A_\mu\,,
\label{loopOp}
\ee
where \(C\) is a closed, light-like, polygonal contour, and the trace is taken over the fundamental representation of the gauge group. The expectation values of such operators are of much interest given their duality with planar scattering amplitudes \cite{Alday:2007hr,Drummond:2007aua,Brandhuber:2007yx,Drummond:2007cf,Bern:2008ap,Drummond:2008aq,Drummond:2007bm}.  

The expectation values of Wilson loop operators enjoy a particularly simple product structure \cite{Drummond:2007cf,Drummond:2007bm,DV1,KK1,Bassetto:1993xd,Drummond:2007au},
\be
W_n = \langle \mathcal{L}(C) \rangle = \Bigl[\prod_{i=1}^n D_i \Bigr] F_n R_n\,.
\label{Wnfactors}
\ee
Here we have a UV divergent factor \(D_i\) associated to each cusp \(x_i\) of the Wilson loop, and split the remaining finite part up such that \(F_n\) fully captures the $O(g^2)$ result, i.e.
\be
R_n = 1 + O(g^4).
\ee
The factor $F_n$ obeys an anomalous conformal Ward identity \cite{Drummond:2007cf,Drummond:2007au} while the factor $R_n$ is conformally invariant. Note that we have defined $R_n$ as an overall finite factor but in many references (e.g. \cite{Dixon:2011pw,Dixon:2013eka,Dixon:2014voa,Drummond:2014ffa}) it is the logarithm of $R_n$ which is referred to as the `remainder function' and this logarithm is non-zero only at two-loop order and beyond.

The objects which we consider here are the correlation functions of \emph{multiple} light-like Wilson loops, which enjoy an analogous factorisation
\be
W_{n_1,\ldots,n_m} = \langle \mathcal{L}(C_1) \ldots \mathcal{L}(C_m)\rangle = \prod_{r=1}^m \biggl[\Bigl[\prod_{i=1_r}^{n_r} D_{i_r} \Bigr] F_{n_r} \biggr]R_{n_1,\ldots,n_m}\,.
\label{Wnmfactors}
\ee
In (\ref{Wnmfactors}), the factors $D_{i_r}$ and $F_{n_r}$ are precisely the same as those appearing in (\ref{Wnfactors}) for a single Wilson loop. The conformally invariant remainder $R_{n_1,\ldots,n_m}$ depends in a complicated way on the kinematic data of all of the Wilson loops which enter into the correlator.

At leading order in \(N\), the dominant contribution to the correlator is of course disconnected contribution, which is simply the product of the individual Wilson loop expectation values. We define the connected part as the remaining contribution. For instance, for the correlator of two loop operators we have 
\be
W_{n_1,n_2} = W_{n_1} W_{n_2} +W_{n_1,n_2}^{\rm conn}\,.
\ee
Since the novel data is entirely contained in the connected part (which is suppressed in by \(1/N^2\) in the large \(N\) limit), this is the piece of the answer which we will focus our attention on. Note that the connected part individually also obeys
\be
W_{n_1,n_2}^{\rm conn} = \prod_{r=1}^2 \biggl[\Bigl[\prod_{i_r=1}^{n_r} D_{i_r} \Bigr] F_{n_r} \biggr]R_{n_1,n_2}^{\rm conn}\,.
\label{connectedFactorisation}
\ee
and we have the same decomposition of the remainder \(R\) into connected and disconnected parts, 
\be
R_{n_1,n_2} = R_{n_1} R_{n_2} + R_{n_1,n_2}^{\rm conn}\,.
\ee

In \cite{Drummond:2025ulh}, we calculated the leading order contributions in \(g^2\) to the connected part in the $SU(N)$ theories, which are UV finite and can be obtained in factorised form where the factors are contributions to the Abelian theory. We refer the reader to \cite{Drummond:2025ulh} for more detail on these calculations and e.g. the colour structure of these objects. At higher order in the coupling, these correlators will clearly develop UV divergences and the loop integrals become substantially more challenging. As such, the \(\bar{Q}\)-equation may provide a vital tool in obtaining results at higher loop order, as has been the case already for a single Wilson loop. 
\subsection{Super Wilson loops}
\subsubsection{\(\mathcal{N}=4\) super Yang-Mills theory in twistor space}
In order to extend the duality between scattering amplitudes and Wilson loop expectation values beyond the MHV sector, it is necessary to introduce a supersymmetric extension of the Wilson loop. There are different but equivalent formulations which appear in the literature, e.g. the spacetime formulation of Caron-Huot \cite{Caron-Huot:2010ryg}, and the twistor Wilson loop of Mason and Skinner \cite{Mason:2010yk}. We make use of the latter here. 

Recall that, perturbatively, \(\mathcal{N}=4\) Super-Yang Mills theory may be captured in supertwistor space using the twistor action \cite{Boels:2006ir}
\be
S(\mathcal{A}) = S_1(\mathcal{A}) + S_2(\mathcal{A}) \,,
\ee
where 
\be
S_1(\mathcal{A}) = \frac{i}{2\pi} \alpha \int D^{3|4} \mathcal{Z} \wedge \textrm{tr} \Bigl(\mathcal{A} \wedge \overline{\partial}\mathcal{A} + \frac{2}{3}\mathcal{A} \wedge \mathcal{A} \wedge \mathcal{A}\Bigr)
\label{selfdual}
\ee
is the action of holomorphic Chern-Simons theory, equivalent to the self-dual sector of the theory, and 
\be
S_2(\mathcal{A}) = g^2_{\rm YM} \alpha^2 \int d^{4|8} X \log \det\bigl(\overline{\partial} + \mathcal{A} \bigr)_X
\label{interaction}
\ee
provides an expansion about the self-dual sector to reproduce the full theory. Here, \(\alpha\) is a free choice of scaling, essentially equivalent to rescaling the fermionic parts of the supertwistors, and resulting from the fact that one must integrate out an auxiliary field in order to convert the twistor action to the ordinary space-time action. This scaling does not affect the overall action but will affect e.g. the resulting expression for the propagator \(\langle \mathcal{A}_a \mathcal{A}_b\rangle\) and thus the perturbative expansion of loop operator expectation values and correlators. In \cite{Drummond:2025ulh}, we took \(\alpha = \frac{N}{4\pi^2}\) in order to faciliate a match between planar Wilson loop expectation values and planar scattering amplitudes beyond the MHV sector. Here, recalling the definition \(C_F = \frac{N^2-1}{2N}\), we slightly refine this choice and choose \(\alpha = \frac{C_F}{2\pi^2}\) (which of course agrees in the large \(N\) limit) in order to ensure the ordinary \(\bar{Q}\) equation for a single Wilson loop holds beyond the planar sector. We will review these non-planar checks of the \(\bar{Q}\) equation later, in Section \ref{Sec-nonPlanarQbar}. Having made this scaling choice we have 
\be
S_1(\mathcal{A}) = \frac{iC_F}{4\pi^3} \int D^{3|4} \mathcal{Z} \wedge \textrm{tr} \Bigl(\mathcal{A} \wedge \overline{\partial}\mathcal{A} + \frac{2}{3}\mathcal{A} \wedge \mathcal{A} \wedge \mathcal{A}\Bigr)
\ee
and
\be
S_2(\mathcal{A}) = \frac{4g^2C_F^2}{\pi^2N}  \int d^{4|8} X \log \det\bigl(\overline{\partial} + \mathcal{A} \bigr)_X\,.
\ee
Note that the pre-factors of \(S_1\)
and \(S_2\) reduce to \(\frac{i}{8\pi^2}\) and \(\frac{g^2N}{\pi^2}\) in the large \(N\) limit. 

For the purpose of perturbative calculations, it is convenient to expand the log-det as a power series in \(\mathcal{A}\). After discarding those terms which do not survive the fermionic integration we are left with 
\be
S_2(\mathcal{A}) = -\frac{4g^2C_F^2}{\pi^2 N} \int d^{4|8}X \sum_{r=2}^{\infty}\frac{1}{r}  
\tr (-\bar{\partial}_X^{-1} \mathcal{A})^r\,.
\ee
\label{S2}
The operator $\bar{\partial}_X^{-1}$ acts on $(0,1)$-forms on the line $X$  as
\be
(\bar{\partial}_X^{-1} \omega)(s) = \int_{X} G(s,s') \wedge \omega(s')\,,
\ee
where we have the Green's function
\be
G(s,s') = -\frac{1}{2\pi i} \frac{ds'}{(s-s')}\,.
\ee
With these definitions we see that
\be
\tr (\bar{\partial}^{-1} \mathcal{A})^r = \tr \biggl\{ \int_{X^r} G(s_r,s_1) \wedge \mathcal{A}(Z(s_1)) \ldots  G(s_{n-1},s_r) \wedge \mathcal{A}(Z(s_r)) \biggr\}\,.
\ee

\subsubsection{The supersymmetric Wilson loop}
In twistor space, the contour for a Wilson loop can be represented by a sequence of intersecting $\mathbb{CP}_1$ lines $X_i$. The intersection points of $X_{i-1}$ and $X_i$ are given by the twistor $Z_i$. We parametrise the lines as 
\be
Z_i(s) = s Z_{i-1} + Z_i\,,
\ee
so that $Z_i(0) = Z_i$ and $Z_i(\infty) = Z_{i-1}$.

The ordinary bosonic loop operator is then given in twistor space as
\be
\mathcal{L} (C)=  \frac{1}{N} \tr \mathcal{P} \prod_{i=1}^n \sum_{l_i=0}^{\infty} \bigl(-\bar{\partial}_i^{-1} a(Z_i(0))\bigr)^{l_i}\,.
\ee
This may be naturally supersymmetrised by simply promoting \(a\) to the full superfield \(\mathcal{A}\),
\be
\mathcal{L} (C)=  \frac{1}{N} \tr \mathcal{P} \prod_{i=1}^n \sum_{l_i=0}^{\infty} \bigl(-\bar{\partial}_i^{-1} \mathcal{A}(Z_i(0))\bigr)^{l_i}\,.
\ee

With our choice of relative scaling for \(S_1\) and \(S_2\), the  propagator is given by 
\begin{align}
\langle \mathcal{A}_a(Z_i(s)) \mathcal{A}_b(Z_j(t)) \rangle^{\rm CS} &=  -\frac{4 \pi^2}{C_F}\delta_{ab} \Delta_*\bigl(Z_i(s),Z_j(t)\bigr) \notag \\ 
&= -\frac{4 \pi^2}{C_F} \delta_{ab} \bar{\delta}^{2|4}\bigl(Z_* , Z_i(s) , Z_j(t)\bigr)\notag \\
&= -\frac{4 \pi^2}{C_F} \delta_{ab} \int \frac{D^2c}{c_1 c_2 c_3} \bar{\delta}^{4|4}\bigl(c_1 Z_* + c_2 Z_i(s) + c_3 Z_j(t)\bigr)\,.
\end{align}
Here, \(\mathcal{Z}_*\) is an arbitrary reference supertwistor which arises as a consequence of having fixed an axial gauge, and upon which well-defined observables should not ultimately depend (although term-by-term they may appear to). Note that the pre-factor reduces to \(-\frac{8\pi^2}{N} \delta_{ab}\) in the large \(N\) limit.

Setting the coupling to zero corresponds to performing the computation in pure Chern-Simons theory. In the large \(N\) limit and for a single Wilson loop, this reproduces tree-level \(\mathcal{N}=4\) scattering amplitudes beyond the MHV sector, with the contribution at order \(4k\) in the Grassmann variables matching the N\(^k\)MHV tree level amplitude. Performing such computations for correlators of multiple loop operators, we similarly have an expansion,
\be
\mathcal{W}^{\rm CS}_{n_1,\ldots,n_m} = \mathcal{W}^{(0), \rm CS}_{n_1,\ldots,n_m} + \mathcal{W}^{(1), \rm CS}_{n_1,\ldots,n_m} + \ldots
\ee
The first term is of Grassmann degree zero and is in fact equal to \(1\), while each subsequent term has Grassmann degree four greater than the last. Although these correlators are no longer obviously dual to amplitudes, we retain the nomenclature e.g. `N$^k$MHV contribution' for the contribution which is of order \(4k\) in the Grassmann variables. Moreover, we continue to refer to the $O(g^0)$ contribution as `tree level' but will avoid e.g. the phrase `one-loop' to avoid conflation with the number of loop operators in the correlator. Note that in the $SU(N)$ theory, the leading non-zero contributions to the fully connected part for $m$ loop operators in the N${}^k$MHV sector are $O(g^{2(m-k)})$ for $k \leq m$ \cite{Drummond:2025ulh}.

\section{Integrands at $O(g^2)$}

\subsection{Loop integrands from twistor diagrams}
Calculations of multiple super Wilson loop correlators in the planar limit at tree level were presented in detail in \cite{Drummond:2025ulh}, and here we turn our attention to the $O(g^2)$ contributions, for which the interaction term \(S_2\) in the action must be included. We have
\be
\langle \mathcal{L}(C_1)\mathcal{L}(C_2) \ldots \mathcal{L}(C_m) \rangle = \int [d\mathcal{A}] e^{- (S_1 + S_2)} \mathcal{L}(C_1)\mathcal{L}(C_2)\ldots\mathcal{L}(C_m)\,.
\ee
which may be perturbatively expanded in the coupling as
\be
\langle \mathcal{L}(C_1)\mathcal{L}(C_2) \ldots \mathcal{L}(C_m)\rangle = \int [d\mathcal{A}] e^{-S_1} \mathcal{L}(C) \biggl[1 + \frac{g^2_{\rm YM}C_F^2}{4\pi^4} \int d^{4|8}X \sum_{r=2}^\infty \frac{1}{r} \tr (-\bar{\partial}_X^{-1} \mathcal{A})^r + O(g_{\rm YM}^4) \biggr]
\ee
The contribution at O($g^2_{\rm}$) is thus given by
\be
\frac{4g^2C_F^2}{\pi^2N} \sum_{r=2}^\infty \frac{1}{r} \int d^{4|8}X \langle \mathcal{L}(C_1)\mathcal{L}(C_2) \ldots \mathcal{L}(C_m)\tr (-\bar{\partial}_X^{-1} \mathcal{A})^r \rangle^{\rm CS} + O(g^4)\,.
\label{gYMsqterm}
\ee

Computing the loop integrand at $O(g^2)$ thus simply amounts to computing the correlator of the Wilson loops with a Lagrangian line \(X\); the loop integration is then simply the integration over the position of the Lagrangian \(X\) in supertwistor space. To compute loop integrands at higher power in the coupling, we would need only include more Lagrangian lines in the correlator.  

For the convenience of the reader, let us collect the rules which allow us to write down the integral corresponding to a given $O(g^2)$ twistor Wilson loop diagram. At N\(^k\)MHV, we receive contributions from diagrams where at least two propagators end on the Lagrangian line, and there are \(k+2\) propagators overall (note that each propagator comes with Grassmann degree \(4\) and the fermionic integration removes \(8\) powers of Grassmann variables). As discussed in \cite{Drummond:2025ulh}, we use a prescription such that diagrams with propagators running between adjacent twistor lines evaluate to zero, and moreover diagrams with two or more propagators running between the same pair of twistor lines will clearly vanish for Grassmann reasons. Such diagrams are therefore not included in our expansion. Note also that when computing the connected part of a correlator of multiple Wilson loops, we disregard disconnected diagrams. 
\begin{itemize}
\item For the Lagrangian line \(X_{AB}\) with \(n \geq 2\) propagator insertions, we include 
\be
\frac{4g^2C_F^2}{N}\int \frac{d^{4|8}x_{AB}}{\pi^2} \biggl(\prod_{p=1}^n \int ds_{x,p} \biggr) \frac{1}{(s_{x,1}-s_{x,2})(s_{x,2}-s_{x,3})...(s_{x,n}-s_{x,1})}.
\ee
Note that diagrams which correspond to cycling the order of the propagator insertions on the Lagrangian are identical, and so we will simply take one of each cyclic class and thus omit the factor \(\frac{1}{r}\) from \(S_2\)'s expansion. The integration variables \(s_{x,1}\) to \(s_{x,n}\) are associated to the insertion points of the propagators on the Lagrangian line \(X_{AB}\), with \(s_{x,1}\) closest to \(\mathcal{Z}_B\) and \(s_{x,n}\) closest to \(\mathcal{Z}_A\). 

\item For external twistor line \(X_i\) on which \(n \geq 1\) propagators end, we include 
\be
\int \frac{ds_{i,1}}{s_{i,1}} \int \frac{ds_{i,2}}{s_{i,2}-s_{i,1}} \int  \frac{ds_{i,3}}{s_{i,3}-s_{i,2}}\, ...  \int  \frac{ds_{i,n}}{s_{i,n}-s_{i,n-1}}.
\ee
Here, the variables \(s_{i,k}\) correspond to the ends of the propagators which end on the twistor line, where \(s_{i,1}\) is closest to \(Z_{i}\) and \(s_{i,n}\) is closest to \(Z_{i-1}\).

\item To the propagator which runs from \((X_{i-1} X_i)\) to \((X_{j-1} X_j)\) (including the possibility that one of the lines is a Lagrangian line), and where the insertion points on the twistor lines are associated to integration variables \(s_{i,m_1}\) and \(s_{j,m_2}\), we associate
\be
\frac{1}{C_F}\int \frac{D^4a_{(i,j)}}{a_{(i,j),1}a_{(i,j),2}a_{(i,j),3}} \bar{\delta}^{4|8}\bigl(a_{(i,j),1}\mathcal{Z}_* + a_{(i,j),2} s_{i,m_1} \mathcal{Z}_{i-1} + a_{(i,j),2}\mathcal{Z}_i + a_{(i,j),3} s_{j,m_2} \mathcal{Z}_{j-1} + a_{(i,j),3}\mathcal{Z}_j\bigr)
\ee

In this rule aside from the normalisation of the propagator itself we have also absorbed a factor of \(-\frac{1}{4\pi^2}\) which will arise from the expansion of \(\bar{\partial}_i^{-1}a(\mathcal{Z}_i(0))\) about each edge on which a propagator is inserted. Note that in the planar limit we have \(\frac{1}{C_F} \to \frac{2}{N}\). Of course, each diagram must be dressed with the appropriate factor coming from the colour trace\footnote{Note that we normalise the gauge group generators as $
\tr (t^a t^b) = \frac{1}{2} \delta_{ab}$}, and the factors of \(2\) from each additional propagator cancels out with an additional power of \(\frac{1}{2}\) on the leading term in the colour trace each time an additional propagator is added to a diagram.

\item For the correlator of \(m\) Wilson loops, include a factor of \(\frac{1}{N^m}\) to account for the normalisations of the loop operators. 
\end{itemize}

Note that, for the $SU(N)$ theory, the $O(g^2)$ connected contribution for $m$ loop operators at N${}^k$MHV is in fact zero for $k \leq m-2$, due to the tracelessness of the generators. E.g. for two loop operators the MHV contribution vanishes at $O(g^2)$. At NMHV, the connected $O(g^2)$ result for two loop operators factorises into the connected NMHV tree multiplied by the connected MHV Abelian $O(g^2)$ contribution which is UV finite and may be directly integrated as in \cite{Drummond:2025ulh}. The first interesting case at $O(g^2)$ for two loop operators is therefore N$^2$MHV which receives contributions from diagrams with four propagators. 

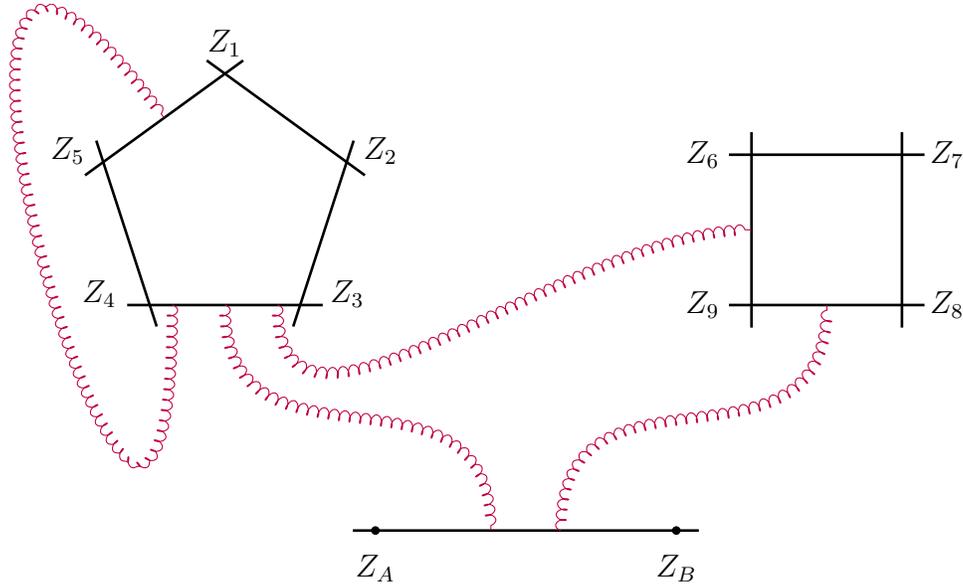
\begin{figure}
\begin{center}
\   \begin{tikzpicture}[baseline={([yshift=-.5ex]current bounding box.center)}]

  \clip (-3.25,-3.75) rectangle (10.75, 4);
  \defineline{lin1}{originx=0.809, originy=2.490, angle=-36, pointcount=3}
  \defineline{lin2}{originx=1+0.309, originy=0.951, angle=72, pointcount=3}
  \defineline{lin3}{originx=0, originy=0, angle=0, pointcount=3}
  \defineline{lin4}{originx=-1-0.309, originy=0.951, angle=-72, pointcount=3}
  \defineline{lin5}{originx=-0.809, originy=2.490, angle=216, pointcount=3}
  
  \drawline[wingorientation=outer,showtwistors=true,tolabel=$\hspace{-6mm}Z_{1}$]{lin1}
  \drawline[wingorientation=outer,showtwistors=true,fromlabel=$\hspace{18.5mm}Z_{2}$,tolabel=$\hspace{22mm}Z_{3}$]{lin2}
  \drawline[showtwistors=true,wingorientation=outer]{lin3}
  \drawline[wingorientation=outer,showtwistors=true,fromlabel=$\hspace{-23mm}Z_{4}$,tolabel=$\hspace{-19mm}Z_{5}$]{lin4}
  \drawline[wingorientation=outer,showtwistors=true]{lin5}

  \defineline{linA}{originx=4, originy=-3, angle=180, pointcount=4,linelength=2}
  \drawline[showtwistors=true,showvertex=true,wingorientation=outer,tolabel=$Z_B$,fromlabel=$Z_A$]{linA}

  \defineline{lin8}{originx=8, originy=0, angle=0, pointcount=3}
  \defineline{lin9}{originx=7, originy=1, angle=270, pointcount=3}
  \defineline{lin10}{originx=8, originy=2, angle=180, pointcount=3}
  \defineline{lin7}{originx=9, originy=1, angle=270, pointcount=3}
  \drawline[showtwistors=true,wingorientation=outer]{lin8}
  \drawline[showtwistors=true,wingorientation=outer,fromlabel=$\hspace{-23mm}Z_{9}$,tolabel=$\hspace{-23mm}Z_{6}$]{lin9}
  \drawline[showtwistors=true,wingorientation=outer]{lin10}
  \drawline[showtwistors=true,wingorientation=outer,tolabel=$\hspace{2mm}Z_{7}$,fromlabel=$\hspace{2mm}Z_{8}$]{lin7}

   \drawpropagator{fromline=lin3, toline=lin5, fromip=1, toip=2,looseval=7.9}
  \drawpropagator{fromline=lin3, toline=lin9, fromip=3, toip=2,looseval=1}
  \drawpropagator{fromline=lin3, toline=linA, fromip=2, toip=3}
  \drawpropagator{fromline=linA, toline=lin8, fromip=2, toip=2}
           
       \end{tikzpicture}

    \end{center}
    \caption{A twistor diagram which contributes at N\({}^2\)MHV to a pentagon-square correlator at $O(g^2)$. Note that this diagram is planar, as can be seen heuristically by the fact that, drawing the Lagrangian line beneath the two Wilson loops and drawing all propagators outside the Wilson loops, it is possible to draw all of the propagators outside the Wilson loops without any crossing.}
\label{oneLoopExample}
\end{figure}

For example, let us consider the diagram depicted in Fig. \ref{oneLoopExample}, which would contribute to the connected part of a correlator of a pentagonal and square Wilson loop, at $O(g^2)$ and N\(^2\)MHV. From our rules, we have the expression (here we relabel some integration variables for notational brevity, for instance writing \(\rho_i\) rather than \(s_{x,i}\) for the integration variables associated to insertions on the Lagrangian line)
\begin{align}
\frac{4g^2}{N^3C_F^2} \textrm{tr}(t_at_bt_ct_c)\textrm{tr}(t_at_d)\textrm{tr}(t_bt_d) \int & \frac{d^{4|8}x_{AB}}{\pi^2} \frac{D^4a}{a_1a_2a_3} \frac{D^4b}{b_1b_2b_3} \frac{D^4c}{c_1c_2c_3} \frac{D^4d}{d_1d_2d_3} \notag \\
& \times \frac{d\rho_1 d\rho_2}{(\rho_1-\rho_2)(\rho_2-\rho_1)} \frac{ds_1}{s_1}\frac{ds_2}{s_2-s_1} \frac{ds_3}{s_3-s_2} \frac{du}{u} \frac{dv}{v} \frac{dw}{w} \notag \\
&\times \bar{\delta}^{4|4}(a_1\mathcal{Z}_* + a_2s_1Z_{3} + a_2\mathcal{Z}_{4} + a_3u\mathcal{Z}_{5} + a_3\mathcal{Z}_{1} ) \notag \\
&\times \bar{\delta}^{4|4}(b_1\mathcal{Z}_* + b_2 s_2 \mathcal{Z}_3 + b_2 \mathcal{Z}_4 + b_3 \rho_2 \mathcal{Z}_{A} + b_3 \mathcal{Z}_{B}) \notag \\
&\times \bar{\delta}^{4|4}(c_1 \mathcal{Z}_* + c_2 s_3 \mathcal{Z}_{3} + c_2 \mathcal{Z}_{4} + v c_3 \mathcal{Z}_9 + c_3 \mathcal{Z}_6) \notag \\
&\times \bar{\delta}^{4|4}(d_1 \mathcal{Z}_* + d_2\rho_1 \mathcal{Z}_{A} + d_2 \mathcal{Z}_{B} + d_3w \mathcal{Z}_{8} + d_3 \mathcal{Z}_{9}) 
\end{align}
Computing the colour trace in the $SU(N)$ theory, the overall pre-factor of the integral is 
\be
g^2\frac{1}{N^2}.
\ee
Note that, as expected for the leading contribution to the connected part, this is \(\frac{1}{N^2}\) suppressed.
This diagram evaluates to give contribution
\begin{equation}
 \frac{g^2}{N^2} \int \frac{d^{4|8}}{\pi^2} [*,5,1,3,4][*,3,\widehat{4}_1,A,\widehat{B}_9][*,3,\widehat{4}_B,9,6][*,A,\widehat{B}_4,8,9]
 \label{examplediag}
\end{equation}
where we define shifted super-twistors
\be
\mathcal{Z}_{i,j} = (i-1 \, i) \cap (j-1 \, j \, *) = \mathcal{Z}_{i-1} \langle i\, j-1\, j\, *\rangle - \mathcal{Z}_{i} \langle i-1\, j-1\, j\, * \rangle .
\ee
The expression (\ref{examplediag}) can be obtained either by directly integrating out on the support of the bosonic delta functions, or by bypassing this procedure using the Feynman rules which we reviewed in \cite{Drummond:2025ulh}. Note that the loop integration remains to be performed and the expression is formal in the sense that in general the contribution from a given diagram will be divergent and in need of regularisation. The fermionic integration is of course simply the instruction to take only those \(\chi\) monomials in the expansion which have maximal dependence on \(\chi_{A}\) and \(\chi_B\) (and specifically those monomials stripped of \(\chi_A\) and \(\chi_B\)), and a useful sanity check on the answer is that those components of the loop integrand which do have maximal dependence on \(\chi_A\) and \(\chi_B\) are independent (both bosonically and fermionically) of the reference supertwistor \(\mathcal{Z}_*\). Note that components which do not survive the fermionic integration can and generally do retain dependence on the reference supertwistor. 

It is straightforward to automate the generation of these diagrams, the computation of their colour factors, and their evaluation e.g. in terms of $R$-invariants as above. Loop integrands at $O(g^2)$ (and indeed higher order in the coupling) may thus be generated without any obstruction for a correlator of any number of Wilson loops. 
\subsection{Finite remainder functions}

As in the case of the expectation value of a single Wilson loop, the $O(g^2)$ contribution to the connected part of the correlator of multiple Wilson loops will in general be UV-divergent, starting at N$^2$MHV for two Wilson loops, and with the divergences factorising over the corners on each loop as given in (\ref{Wnmfactors}). The finite, conformally invariant factor $\mathcal{R}$ will be of particular interest since it is what enters into the \(\bar{Q}\)-equation. At fixed MHV degree, we may express the finite part of the $O(g^2)$ contribution to the connected correlator as 
\begin{equation}
 \mathcal{R}^{(k,1), \textrm{conn} }_{n_1,\ldots,n_m} = \mathcal{W}^{(k,1), \textrm{conn}}_{n_1,\ldots,n_m} - \mathcal{W}^{(k,0), \textrm{conn}}_{n_1,\ldots,n_m} \sum_{r=1}^m \mathcal{W}^{(0,1)}_{n_r}
 \label{connectedremainder}
\end{equation}
i.e. we subtract the tree-level connected part at the same MHV degree, multiplied by the sum of the one-loop MHV contributions to each individual Wilson loop. Here for notational brevity we write superscript \((k,l)\) to denote the contribution at N\({}^k\)MHV and \(O(g^{2l})\).
\subsection{Comment on integrands involving triangular Wilson loops}
At MHV level we have already noted in \cite{Drummond:2025ulh} that correlators of two Wilson loops where one is a triangle will always vanish in real kinematics; when both Wilson loops are triangles, the function vanishes even in complex kinematics. The same observations obviously carry over to NMHV given the factorisation of the NHMV one-loop result into the NMHV tree level contribution times the MHV one-loop contribution, even in the $SU(N)$ theory.

In real kinematics, it is straightforward to check that correlators involving triangles are zero \emph{even at the level of the integrand} for all MHV degree; this is the case even for each twistor diagram individually. We may impose real kinematics at the level of supertwistors by shrinking all of the supertwistors encoding the kinematics of that Wilson loop to a point, i.e. taking
\begin{equation}
\mathcal{Z}_2 = \mathcal{Z}_1 + \epsilon(\mathcal{Z}_{r_1} + a\mathcal{Z}_{r_2} + bZ_{r_3})
\end{equation}
\begin{equation}
\mathcal{Z}_3 = \mathcal{Z}_1 + \epsilon(\mathcal{Z}_{r_1} + c\mathcal{Z}_{r_2} + d\mathcal{Z}_{r_3})
\end{equation}
and taking the \(\epsilon \to 0\) limit. Note that here \(Z_{r_i}\) are three supertwistors other than the three \(\mathcal{Z}_1\), \(\mathcal{Z}_2\), \(\mathcal{Z}_3\)  defining the triangle, which can e.g. be taken from the other Wilson loop. It is then straightforward  to check that in this limit each diagram is indeed \(\epsilon\)-suppressed.

However, the integrands are in general non-zero in complex kinematics even when both Wilson loops are triangles.

\section{Local expressions for the integrand and integrated correlator}
\subsection{A review of the chiral box expansion}

The $O(g^2)$ expression for a Wilson loop's expectation value, or a correlator of multiple Wilson loops, is in general divergent and in need of regularisation; for instance, using the dual conformal regularisation scheme introduced in \cite{Bourjaily:2013mma}. In the same reference, the \emph{chiral box expansion}, a basis of manifestly local one-loop integrals, was introduced. Since all of these integrals have a simple and known analytic answer, the first (and only non-trivial) step towards converting the loop integrand to the integrated answer is to rewrite the loop integrand in terms of this chiral box basis. 

Recall the elementary result from Schubert calculus that, in general, there are two distinct configurations (\emph{Schubert solutions}) of the twistor line \((AB)\) which intersect four given, sufficiently generic lines (the \emph{Schubert problem}). This means that, when considering the residue on a given maximal cut e.g. the configuration
\[\langle AB 12 \rangle = \langle AB 23 \rangle = \langle AB45 \rangle = \langle AB56 \rangle = 0,\]
we must specify which Schubert solution we are actually taking the residue on. The key idea of the chiral box basis is that for each given physical maximal cut (i.e. a choice of four twistor brackets of the form \(\langle AB\,i-1,i \rangle\) to set to zero), there are \emph{two} chiral box integrands, each associated to one or the other Schubert solution. Each chiral box integrand has residue\footnote{Strictly speaking the residue is \(\pm\frac{1}{\pi^2}\) once the integration measure is properly normalised, but as is standard in the literature we will nonetheless refer to this as 'unit leading singularity', and use the shorthand \(\pm1\) for these residues throughout the rest of our discussion.} \(\pm 1\) on one Schubert solution, and \(0\) on the other. By dressing each chiral box integrand with plus or minus the appropriate residue on the appropriate Schubert solution, we may therefore match all physical maximal cuts of a linear combination of chiral box integrands to our original loop integrand (e.g. computed via diagrams).

We collect expressions for each chiral box integrand, and its integrated value, in Appendix \ref{chiralBoxes}. Note that while each chiral box integrand/integral in general depends on an auxiliary bi-twistor \(X \equiv (X_1X_2)\), this cancels out in the overall expression, which provides a useful sanity check on the answer. It is important to note that each chiral box integrand has been carefully constructed so as to have both the correct residues on the Schubert solutions for the appropriate physical quadruple pole, \emph{and also zero residue on four-mass type cuts involving the auxiliary \(X\)}. For instance,
\be
\langle AB12 \rangle = \langle AB34 \rangle = \langle AB56 \rangle = \langle ABX \rangle = 0
\ee
would be an example of such a cut which is possible for hexagons and higher multiplicity Wilson loops. As such, it is not the case that e.g. the three mass integrands/integrals follow from a collinear limit on the four mass integrands/integrals. Note that two-mass hard cuts involving \(X\) such as 
\be
\langle AB12 \rangle = \langle AB23 \rangle = \langle AB34 \rangle = \langle ABX \rangle
\ee
still give non-zero residue on some of the chiral box integrands, and it is the need to cancel these from the overall answer which allows us to fix the coefficients of the so-called 'triangle' integrals which capture the divergences and which we will discuss shortly.
\subsubsection*{Example: one-mass chiral boxes}
For example, consider the expectation value of a single Wilson loop with \(n > 4\) sides. An example of a one-mass physical quadruple pole would correspond to the configuration when
\begin{equation}
\langle AB12 \rangle = \langle AB23 \rangle = \langle AB34 \rangle = \langle AB45 \rangle = 0. 
\end{equation}
As is well-known (see e.g. \cite{Arkani-Hamed:2010pyv} for a review of the solutions to the various Schubert problems), there are two solutions to this Schubert problem, i.e. two configurations of the line \(AB\) such that the four twistor brackets simultaneously vanish. In particular, since the vanishing of \(\langle A \, B \, i-1 \, i\rangle \) corresponds to the requirement that the lines \((AB)\) and \((i-1  \ i)\) intersect, we can choose 
\begin{equation}
(AB) = (24) \quad \text{or} \quad 
(AB) = (123) \cap (345).
\end{equation}
The first solution is obvious: \(Z_2\) lies on both \((12)\) and \((23)\), while \(Z_4\) lies on both \((34)\) and \((45)\), which means the line \((24)\) clearly cuts all of the lines \((12)\), \((23)\), \((34)\) and \((45)\). 

The second solution is similarly apparent: any line in the plane \((123)\) will certainly intersect the lines \((12)\) and \((23)\), while any line in the plane \((345)\) will intersect \((34)\) and \((45)\). So, by choosing \((AB)\) to the intersection of these planes we again find that we successfully intersect all four lines. 

The chiral box expansion encompasses two separate integrands associated to this physical quad-cut; one for the first Schubert solution, and one for the second. Labelling the box by its propagators, these integrands are 

\begin{equation}
\chi^{1m}_{(12)(23)(34)(45),1} = \int \frac{d^4x_{AB}}{\pi^2} \frac{\langle AB (123)\cap(345) \rangle \langle X 24 \rangle}{\langle AB12\rangle \langle AB23 \rangle \langle AB34 \rangle \langle AB45 \rangle \langle ABX\rangle} 
\end{equation}
and
\begin{equation}
\chi^{1m}_{(12)(23)(34)(45),2} =  \int \frac{d^4x_{AB}}{\pi^2} \frac{\langle AB24 \rangle \langle X (123) \cap (345) \rangle}{\langle AB12\rangle \langle AB23 \rangle \langle AB34 \rangle \langle AB45 \rangle \langle ABX\rangle}.
\end{equation}

The first integrand has a residue of \(1\) on the Schubert solution \((AB)=(24)\) and \(0\) on the second Schubert solution \((123)\cap(345)\); the second integrand has a residue of \(0\) on the first Schubert solution and \(-1\) on the second. Clearly there are no further physical quadruple poles. As such, by dressing the first integrand with the residue of the observable computed on the first Schubert solution, and the second integrand with minus the residue of the observable computed in the second Schubert solution, we may ensure that our box expansion matches the residue on this physical pole, and can repeat this procedure to construct a sum of local integrands with the correct residue on every physical pole (separately on each Schubert solution). Note that for a single amplitude/Wilson loop expectation value, compact expressions for these residues are readily available via e.g. the formalism of on-shell diagrams, while at least in the first instance for our correlators we will need to compute the residues explicitly using the expression for the loop integrand computed via twistor Wilson loop diagrams.

\subsubsection*{Example: zero-mass chiral boxes}
The zero-mass chiral box is essentially irrelevant for the study of a single Wilson loop's expectation value, since it would only be relevant in the case \(n=4\). As such, it was omitted in \cite{Bourjaily:2013mma}. However, since we will certainly concern ourselves with correlators of multiple Wilson loops including a square, the zero-mass chiral box will be important in our case. 

It is straightforward to see that the correct choice for the chiral box integrands are (here for simplicity we consider the case of \(\langle AB12 \rangle = \langle AB23 \rangle = \langle AB34 \rangle = \langle AB41 \rangle = 0\))

\begin{equation}
\chi^{0m}_{(12)(23)(34)(41),1} = \int \frac{d^4x_{AB}}{\pi^2}\frac{\langle 1234 \rangle \langle X 24 \rangle \langle AB13 \rangle}{\langle AB12 \rangle \langle AB23 \rangle \langle AB34 \rangle \langle AB41 \rangle \langle ABX \rangle }
\end{equation}
and
\begin{equation}
\chi^{0m}_{(12)(23)(34)(41),2} = \int \frac{d^4x_{AB}}{\pi^2}\frac{\langle 1234 \rangle \langle X 13 \rangle \langle AB24 \rangle}{\langle AB12 \rangle \langle AB23 \rangle \langle AB34 \rangle \langle AB41 \rangle \langle ABX \rangle }
\end{equation}

where the first has residue \(+1\) on the solution \((AB) = (24)\) and \(0\) on \((AB) = (13)\), and the second has residue \(0\) on the solution \((AB) = (24)\) and \(-1\) on \((AB) = (13)\). As for all other chiral boxes, the integrated expressions are identical for the two chiralities, and for this configuration we have integrated expression  
\begin{equation}
-3\Li_2(1)- \frac{1}{2}\log\biggl(\frac{\langle 12X \rangle \langle 34X \rangle}{\langle 23X \rangle \langle 14X \rangle}\biggr)^2.
\end{equation}

\subsubsection*{Example: the four-mass Schubert problem}
Let us briefly review the solution to the four mass Schubert problem, for which the geometry is less transparent and the solutions are more complicated. For instance, consider the four mass Schubert problem
\begin{equation}
\langle AB12 \rangle = \langle AB34 \rangle = \langle AB 56 \rangle = \langle AB78 \rangle = 0
\end{equation}
which would be relevant e.g. for an octagon. To solve the Schubert problem, we seek a configuration of the line \(AB\) which intersects all of the lines \((12)\), \((34)\), \((56)\) and \((78)\). Clearly by choosing \(A\) and \(B\) to lie on the lines \((12)\) and \((56)\) respectively, i.e.
\begin{align}
Z_A &= Z_1 + \alpha Z_2\,, \\
Z_B &= Z_5 + \beta Z_6 \,,
\end{align}
we can ensure that \(\langle AB12 \rangle = \langle AB56 \rangle = 0\), and need only choose \(\alpha\) and \(\beta\) such that the other two determinants also vanish. Plugging our parametrisations into \(\langle AB34 \rangle\) and \(\langle AB78 \rangle\), this amounts to the following pair of conditions
\begin{align}
\langle 1534 \rangle + \alpha \langle 2534 \rangle + \beta \langle 1634 \rangle + \alpha \beta \langle2634 \rangle &= 0 \,, \\
\langle 1578 \rangle + \alpha \langle 2578 \rangle + \beta \langle 1678 \rangle + \alpha \beta \langle2678 \rangle &= 0 \,.
\end{align}
These simultaneous equations in \(\alpha\) and \(\beta\) amount to a quadratic equation in \(\alpha\) or \(\beta\) which may readily be solved; the choice of sign for the square root is what selects the Schubert solution, and so the expression for \(Z_A\) (or \(Z_B\)) on one Schubert solution is the Galois conjugate of the solution for \(Z_A\) (or \(Z_B\)) on the other. 

\subsubsection{Triangle integrands}
By dressing each chiral box with (plus or minus) the correct residue, we can write down a linear combination of local integrands which matches the residues on each Schubert solution for every physical cut of the observable. However, these chiral boxes alone cannot correctly match the answer, since every chiral box is finite and yet in general the one-loop answer should be UV divergent. The missing ingredient is supplied by the divergent \emph{triangle integrals}, which contain only three physical poles. These are of the form
\begin{equation}
\mathcal{I}_{a} = \int \frac{d^4x_{AB}}{\pi^2}\frac{\langle a-1 a a+1 a+2 \rangle \langle X a+1 \, a \rangle}{\langle AB a-1 a \rangle \langle AB a+1 a \rangle \langle AB a+1 a+2 \rangle \langle AB X \rangle}.
\end{equation}
Aside from supplying the divergences which we know are present in the final answer, these integrands also play another crucial role: cancelling out the residues on two-mass hard type cuts which involve the auxilliary twistor \(X\), which the chiral box expansion \emph{sans triangles} would include. Explicitly working out the coefficient each triangle needs to be dressed with in order to cancel out these residues, one finds that for the connected contribution to a multiple Wilson loop correlator it is simply the N\(^k\)MHV tree-level result for that connected correlator, as expected from the divergence structure discussed earlier.

\subsubsection{MHV contributions at $O(g^2)$ and equivalence with the chiral pentagon expansion}
It is instructive to make contact with the familiar story of the \emph{chiral pentagon expansion} for one-loop MHV integrands, originally presented in \cite{Arkani-Hamed:2010pyv}. We define the chiral pentagon integral as
\begin{equation}
P_{ij} = \int \frac{d^4x_{AB}}{\pi^2} \frac{\langle AB (i-1 i i+1) \cap (j-1 j j+1) \rangle \langle AB X \rangle}{\langle i-1 i AB \rangle  \langle i i+1 AB \rangle \langle j-1 j AB \rangle \langle j j+1 A B \rangle \langle AB X \rangle }.
\end{equation}

As is well-known \cite{Arkani-Hamed:2010pyv}, for a single \(n\)-sided Wilson loop the one-loop MHV contribution may be expressed as
\begin{equation}
-\sum_{i < j} P_{ij}
\end{equation}
where \(i\) and \(j\) run over the indices \(1\) to \(n\) and the \(i < j\) condition is to be understood modulo \(n\). Similarly, we showed in \cite{Drummond:2025ulh} that the connected MHV contribution to the correlator of \emph{two} Wilson loops at \(O(g^2)\) is given by 
\begin{equation}
-\sum_{i,j} P_{ij}
\end{equation}
where \(i\) runs over the indices of the first Wilson loop and \(j\) runs over the indices of the second Wilson loop. 

These chiral pentagons may be identified precisely with the appropriate chiral boxes. Namely:
\begin{itemize}
\item Where \(i\) and \(j\) are seperated by more than two, or come from different Wilson loops, \(P_{ij}\) is simply the two-mass easy chiral box on the first Schubert solution (in our labelling convention given in the appendix). So for instance for a single octagon,
\begin{equation}
P_{25} = \chi^{2me}_{(12)(23)(45)(56), 1}\,.
\end{equation}

\item Where \(i\) and \(j\) are separated by two and the associated Wilson loop is at least a pentagon, \(P_{ij}\) is simply the one-mass chiral box on the first Schubert solution (in our labelling convention given in the appendix). So for instance for a single octagon,
\begin{equation}
P_{24} = \chi^{1m}_{(12)(23)(34)(45), 1}\,.
\end{equation}

\item Where \(i\) and \(j\) are separated by two and the associated Wilson loop is a square, \(P_{ij}\) is a zero-mass chiral box, and in fact over the sum we pick up both Schubert solutions. For instance for a square we have
\begin{equation}
P_{24} = \chi^{0m}_{(12)(23)(34)(41), 1} 
\end{equation}
and
\begin{equation}
P_{13} = \chi^{0m}_{(12)(23)(34)(41), 2} \,.
\end{equation}

\item Where \(i\) and \(j\) are separated by one, the chiral pentagon degenerates to precisely minus the divergent triangle integrands. For instance, 
\begin{equation}
P_{23} = \frac{\langle 1234 \rangle \langle 23 X \rangle}{\langle AB12 \rangle \langle AB23 \rangle \langle AB34 \rangle \langle ABX \rangle}  = -\mathcal{I}_1.
\end{equation}

\end{itemize}

Since we have already observed that the coefficient of the divergent triangles is just the N\(^k\)MHV tree-level result, when we construct the remainder function as in eq. (\ref{connectedremainder}) by subtracting the MHV one-loop contribution times the appropriate tree-level result, it is clear that the triangles simply cancel out entirely. The coefficients of the other chiral boxes which enter into the sum over chiral pentagons are simply modified by adding the N\(^k\)MHV tree result to the original coefficient of those integrals (which, prior to converting to the remainder function, was the residue of the correlator/expectation value on that cut). 

\subsubsection{Chiral boxes for multiple Wilson loop correlators}
By simply dressing each chiral box with the appropriate sign-dressed residue of the integrand, we may construct a manifestly local expression for the correlator of multiple light-like Wilson loops, including the triangle integrals which supply the UV divergences and cancel the spurious two-mass hard poles involving the auxiliary point \(X\). Aside from the useful sanity check of independence on the auxiliary point \(X\), it is also easy to test numerically that this manifestly local form of the loop integrand matches the original integrand which was computed diagrammatically.

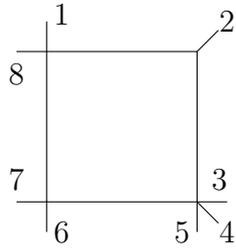
\begin{figure}
\begin{center}
  \begin{tikzpicture}
\draw (0,0) -- (2,0) -- (2,2) -- (0,2) -- cycle;

\draw (0,2) -- (-0.4,2);
\draw (0,2) -- (0,2.4);
\draw (0,0) -- (-0.4,0);
\draw (0,0) -- (0,-0.4);
\draw (2,2) -- (2.283,2.283);
\draw (2,0) -- (2.4,0);
\draw (2,0) -- (2,-0.4);
\draw (2,0) -- (2.283,-0.283);

\node at (-0.4,1.7) {$8$};
\node at (0.2,2.5) {$1$};
\node at (2.4,2.4) {$2$};
\node at (2.3,0.3) {$3$};
\node at (2.4,-0.4) {$4$};
\node at (1.8,-0.4) {$5$};
\node at (0.2,-0.4) {$6$};
\node at (-0.4,0.3) {$7$};
\end{tikzpicture}
\end{center}
\caption{The standard notation for the three-mass quadruple cut corresponding to \(\langle AB12 \rangle = \langle AB23 \rangle = \langle AB56 \rangle = \langle AB78 \rangle = 0\). For a single Wilson loop there is a natural ordering in which to put the legs, corresponding to the cyclic ordering of the Wilson loop itself.}
\label{octagonBox}
\end{figure}

It is worth addressing a subtlety in the definition of the chiral box integrands. In the case of a single Wilson loop, there is a natural cyclic ordering of all of the edges of the Wilson loop, and as such it is customary to label the chiral box integrands by their \emph{legs} and to draw diagrams depicting this natural ordering. For instance, for the three-mass cut \((12)(23)(56)(78)\) for a single octagon, we would usually draw the diagram depicted in Fig. \ref{octagonBox} to represent the cut.

\begin{figure}
\begin{center}
  \begin{tikzpicture}
\draw (-3,0) -- (-1,0) -- (-1,2) -- (-3,2) -- cycle;

\draw (-3,2) -- (-3.4,2);
\draw (-3,2) -- (-3,2.4);
\draw (-3,0) -- (-3.4,0);
\draw (-3,0) -- (-3,-0.4);
\draw (-3,2) -- (-3.283,2.283);
\draw (-1,2) -- (-0.717,2.283);
\draw (-1,0) -- (-0.6,0);
\draw (-1,0) -- (-1,-0.4);
\draw (-1,0) -- (-0.717,-0.283);

\node at (-3.4,1.7) {$8$};
\node at (-2.8,2.5) {$1$};
\node at (-0.6,2.4) {$2$};
\node at (-0.7,0.3) {$3$};
\node at (-0.6,-0.4) {$?$};
\node at (-1.2,-0.4) {$5$};
\node at (-2.8,-0.4) {$6$};
\node at (-3.4,0.3) {$7$};
\node at (-3.4,2.4) {$?$};

\draw (3,0) -- (5,0) -- (5,2) -- (3,2) -- cycle;

\draw (3,2) -- (2.6,2);
\draw (3,2) -- (3,2.4);
\draw (3,0) -- (2.6,0);
\draw (3,0) -- (3,-0.4);
\draw (3,2) -- (2.717,2.283);
\draw (5,2) -- (5.283,2.283);
\draw (5,0) -- (5.4,0);
\draw (5,0) -- (5,-0.4);
\draw (5,0) -- (5.283,-0.283);

\node at (2.6,1.7) {$6$};
\node at (3.2,2.5) {$1$};
\node at (5.4,2.4) {$2$};
\node at (5.3,0.3) {$3$};
\node at (5.4,-0.4) {$?$};
\node at (4.8,-0.4) {$7$};
\node at (3.2,-0.4) {$8$};
\node at (2.6,0.3) {$5$};
\node at (2.6,2.4) {$?$};
\end{tikzpicture}
\end{center}
\caption{Two natural ways of drawing the same three mass cut as in Fig. \ref{octagonBox}, but for a square-square correlator rather than a single octagon. There is no longer a natural ordering for the propagators being cut, nor is there a natural identity for the legs between labels on different Wilson loops.}
\label{newBoxes}
\end{figure}
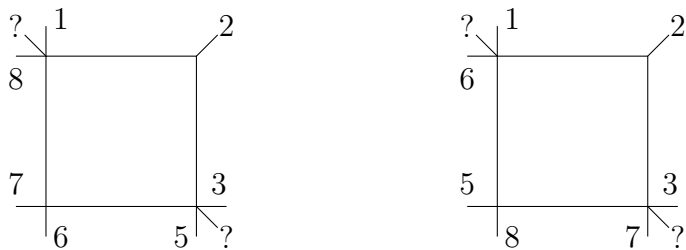

For a correlator involving e.g. two Wilson loops, quadruple cuts involving propagators drawn from both Wilson loops can have an ambiguity in their ordering. For instance, for a square-square correlator we \emph{also} have the three-mass cut \((12)(23)(56)(78)\) but now it is equally natural to depict the box diagram as either of the variants depicted in Fig. \ref{newBoxes}, and there is no longer a natural identification for the legs at the corners when we cross from one Wilson loop to the other. 

To avoid any ambiguity, we will always label the chiral boxes by the \emph{propagators}, not the legs. In this way the chiral box integrands corresponding to the cuts on the left and right would respectively be labelled
\begin{equation}
\chi^{3m}_{(12)(23)(56)(78), i} \, \, \, \, \,  \chi^{3m}_{(12)(23)(78)(56),i}
\label{twoChiralIntegrands}
\end{equation}
for \(i=1\) or \(i=2\) depending on which Schubert solution the box corresponds to. Using the formulae for the appropriate chiral box integrands, it is easy to see that these two orderings differ by a minus sign. This is in fact a pleasing feature. Recall that when computing the multidimensional residue associated to a particular quadruple cut, one implicitly chooses an \emph{ordering} of the four poles to set to zero, and a pairwise swap of two poles in this order will induce a minus sign change due to the swapping of two rows in the Jacobian which one computes. As such, when we make the claim that a given chiral box integrand has a residue of \(1\) or \(-1\) on a particular Schubert solution, we implicitly mean that it has that residue on that Schubert solution \emph{when we put the poles in a particular order during the residue computation}. The difference between the two integrands in (\ref{twoChiralIntegrands})
is that they each have a residue of \(1\) when computing the cut 
\[\langle AB12 \rangle = \langle AB23 \rangle = \langle AB56 \rangle = \langle AB78 \rangle = 0,\] on the first Schubert solution, but for the former the four poles must be put in the order \(\langle AB12 \rangle, \langle AB23 \rangle, \langle AB56 \rangle, \langle AB78 \rangle\) during the residue computation, and for the latter the four poles must be put in the order \(\langle AB12 \rangle, \langle AB23 \rangle, \langle AB78 \rangle\), \(\langle AB56 \rangle\).

When expanding a loop integrand in terms of a sum of chiral boxes, we therefore only use one representative for each choice of four poles, and simply keep track of our order of those poles to avoid a minus sign error when we compute the residue which will dress this chiral box as its coefficient.
\subsubsection{Computation of leading singularities}
\label{compleadingsings}
Since the coefficients of the chiral box integrands are by design precisely (plus or minus) the residues of the one-loop integrand on the corresponding physical quad-cut, it is necessary to study these leading singularities of the integrand. In the case of a single Wilson loop/amplitude, the formalism of on-shell diagrams supplies simple expressions for the leading singularities which are effectively a product of tree-level objects. While we might hope that a similar story holds for our present case, and will indeed present such formulae in a companion paper, in the first instance it is necessary to compute these residues by explicit computation. We find that it is easiest to proceed diagram by diagram and then to sum the overall contribution to the residue at the end. 

We have already noted that each N\({}^k\)MHV diagram can be evaluated as a product of \(k+2\) $R$-invariants. Recalling the explicit form of an $R$-invariant 
\be
[a,b,c,d,e] = \frac{\overline{\delta}^{0|4}(\langle bcde \rangle \chi_{a} + \textrm{cyc})}{\langle abcd \rangle \langle bcde \rangle \langle cdea \rangle \langle deab \rangle \langle eabc \rangle}
\ee
the contribution from a diagram may be written explicitly as 
\begin{equation}
f(Z) \prod_{i=1}^{k+2} \overline{\delta}^{4|4}(\sum_{j=1}^5 r_{i,j}(Z) \mathcal{Z}_{i,j}).
\end{equation}
Here, \(f(Z)\) is a rational, bosonic function in twistor brackets (coming from the denominators associated to each $R$-invariant); \(r_{i,j}(Z)\) are polynomials in twistor brackets giving the coefficient of each Grassmann variable inside the delta functions; and \(\mathcal{Z}_{i,j}\) are supertwistors which may be drawn from the physical twistors specifying the kinematics of the Wilson loops as well as the reference twistor \(\mathcal{Z}_*\), the loop twistors \(\mathcal{Z}_A\) and \(\mathcal{Z}_B\), and shifted supertwistors of the form \((i-1 \, i) \cap (j-1 \, j \, *)\). 

To compute the residue of a given diagram on a particular Schubert solution associated to a particular quadruple cut, we can follow the simple procedure below. While we omit a detailed discussion of the computation of multidimensional residues here, a thorough review in precisely this context is given in \cite{Arkani-Hamed:2010pyv}. 

\begin{enumerate}
\item Since all the poles are isolated in the bosonic factor \(f(Z)\), we can take the cut in the ordinary sense on that term. That is: multiply \(f(Z)\) by the four poles being set to zero; multiply by the appropriate Jacobian corresponding to the multidimensional residue; and plug in for \(Z_A\) and \(Z_B\) using a parametrisation of the appropriate Schubert solution. 
\item Inside the delta functions, replace the bosonic parts of the loop twistors \(Z_A\) and \(Z_B\) with the parameterisation of the solution to the appropriate Schubert problem; leave the fermionic parts \(\chi_A\) and \(\chi_B\) untouched as these are really placeholders for when we perform the fermionic integration and must select terms with maximal dependence on these Grassmann variables.
\item The requirement to perform the fermionic integration on the residue is then interpreted as the instruction to only inspect \(\chi\) monomials with maximal dependence on \(\chi_A\) and \(\chi_B\). 
\end{enumerate}

By carrying out this process for each diagram and summing over the contributions, we are able to compute the residue on a given Schubert solution to a given physical quad-cut, and by the prior discussion we know that it is this residue (dressed with a minus sign for one Schubert solution) which provides the box coefficient in the chiral box expansion of the loop integrand. While the leading singularities computed explicitly in this way will generically appear to be complicated functions, they of course have no dependence on the loop variables; this means the entire dependence on \(Z_A\) and \(Z_B\) is successfully packaged up in the chiral box integrands, whose integrated expressions are known. It is then a straightforward sanity check to numerically sample the local expression for the integrand which is obtained in this way and verify that it really does match the original expression obtained via diagrams. 

The end result of the above process is an expression for the integrand of the finite remainder with the loop line $x_{AB}$ still to be integrated over. Thus we obtain the following form for the reaminder (denoting the connected part by `c'),
\begin{equation}
\mathcal{R}_{n_1,\ldots,n_m}^{(k,1),{\rm c}} = \int d^4x_{AB}\sum_{P} \bigl(C_{P,1}^k \chi_{P,1}(x_{AB}) + C_{P,2}^k \chi_{P,2}(x_{AB})\bigl) \,,
\end{equation}
where the sum is over all finite chiral box integrands labelled by a choice of propagators $P$ and labelled `1' or `2' to distinguish the two chiralities. Since we give the expression for the remainder function rather than the raw correlator, note that `triangle' integrals are absent, and certain chiral box coefficients (namely, both chiralities of zero-mass boxes, as well as the first chirality for one-mass and two-mass boxes on the first chirality) are modified (following eq. (\ref{connectedremainder})) by the addition of $\mathcal{W}^{(k,0),\rm{c}}_{n_1,\ldots,n_m}$. Prior to this modification, \(C^k_{P,1}\) is given by the leading singularity on the appropriate cut and the first Schubert solution, and \(C^k_{P,2}\) is given by \emph{minus} the leading singularity on the second Schubert solution\footnote{This is because, in our conventions, the corresponding chiral box integrand has a residue of \(-1\) rather than \(1\) on the appropriate quadruple cut.}.

Since all of the chiral box integrals are known, once we have expressed the integrand in a local form we are able to immediately write down the integrated answer. The two conjugate chiral boxes for a given configuration of propagators are identical at the integrated level and we obtain for the integrated answer,
\begin{equation}
\mathcal{R}_{n_1,\ldots,n_m}^{(k,1),{\rm c}} = \sum_{P} \bigl(C_{P,1}^k + C_{P,2}^k \bigl)\bar{\chi}_{P} = \sum_{P} C_{P}^k \bar{\chi}_{P} \,,
\end{equation} 
where
\be
\bar{\chi}_P = \int d^4x_{AB}\, \chi_{P,1}(x_{AB}) = \int d^4 x_{AB}\, \chi_{P,2}(x_{AB})
\ee 
is the integrated expression for either chiral box with a given set of four propagators.

In the ancillary files, we present the $O(g^2)$ N\({}^2\)MHV contribution to the correlator of two polygons for square-square, square-pentagon, pentagon-pentagon and the N$^3$MHV contribution to the correlator of two squares. All of these calculations are performed in the planar limit, and we have expressed the box coefficients in terms of $R$-invariants by using the general formulae which we will derive and present in more detail in a companion paper; we have also checked that the chiral box coefficients given match those obtained by a direct, diagrammatic computation of the residues.  

\subsection{Integrated expressions involving trianglular loops}

Recall that, in order to obtain an \emph{integrated} expression for Wilson loop correlators, our approach has been to obtain an expression in terms of the chiral box basis of integrands, whose integrals are known. Obtaining such an expansion essentially amounts to obtaining expressions for the residues on the various physical quad-cuts that the integrand possesses, on each of the two solutions to the relevant Schubert problems. 

Issues arise with the analysis of Schubert problems for correlators involving triangular Wilson loops.  The most obvious issue arises for the quad-cut involving three propagators on the triangle plus one on the other polygon. For instance, if the cusps of the triangle are \(\mathcal{Z}_1\), \(\mathcal{Z}_2\) and \(\mathcal{Z}_3\), such a quad-cut would correspond to 
\begin{equation}
\langle AB12 \rangle = \langle AB23 \rangle = \langle AB31 \rangle = \langle AB i-1 i \rangle = 0.
\end{equation}
where \(\mathcal{Z}_i\) is on the other Wilson loop. Such a Schubert problem is clearly insufficiently generic to possess only two discrete solutions, as there is a continuous one-parameter family of solutions rather than the usual two, corresponding to all lines which pass through the point \((123) \cap (i-1 i)\) and lie in the plane \((123)\). 

There are also issues with certain Schubert solutions setting other poles to zero which, uniquely for triangles, are physical. For instance, consider the N\(^2\)MHV $O(g^2)$ \(\langle\mathcal{W}_{1,2,3}\mathcal{W}_{4,5,6,7} \rangle\) correlator, and consider the two-mass easy quad-cut 
\begin{equation}
\langle AB12 \rangle = \langle AB23 \rangle = \langle AB45 \rangle = \langle AB56 \rangle = 0.
\end{equation}
The second Schubert solution
\begin{equation}
(AB) = (123)\cap (456),
\end{equation}
has the unfortunate effect that \(\langle AB 31 \rangle\) is automatically also set to zero along with the intended four poles. Ordinarily this is completely harmless, since \(\langle AB31 \rangle\) is not a physical pole of any correlator which doesn't involve a triangle, and non-physical poles which don't involve the reference twistor don't even appear at the level of individual diagrams. However, for our correlator involving a triangle, \(\langle AB31 \rangle\) \emph{is} a physical pole which means that a naive computation of the residue on this Schubert solution takes contributions from diagrams on which we have \(\langle AB31 \rangle\) as a pole but \emph{not} e.g. \( \langle AB23 \rangle\). 

Nonetheless, we may still take e.g. our local expression for a square-square correlator and send the square to a triangle via a collinear limit. Doing so, several integrals develop a \(\log(\epsilon)\) divergence. While one might hope that \(\log(\epsilon)\) will cancel out in the overall sum (which does indeed happen when e.g. sending an $(n+1)$-gon to an $n$-gon via collinear limits for $n>3$), this does \emph{not} seem to be the case, and so the limit to an integrated triangle correlator at N\(^2\)MHV appears genuinely divergent in complex kinematics and in need of some sort of additional regularisation. 

However, in real kinematics it is simple to check that, even at the level of the integrand, expressions involving triangles vanish. As already noted, this happens twistor diagram by twistor diagram.

\subsection{Example: N\(^2\)MHV square-square correlator}
As a concrete example, let us consider the N\(^2\)MHV contribution to a square-square correlator at $O(g^2)$, which is simple enough to present explicitly here. Let us emphasise that we do not derive here these relatively compact expressions for the chiral box coefficients (i.e. the leading singularities), but will discuss these in more detail in \cite{leadingSing}. In the below, we recall (here we use a superscript `c' to denote the connected part)
\begin{align}
\mathcal{W}_{I}^{(1,0)} &= \sum_{(i<j) \in I}  [*,i-1,i,j-1,j]\,, \\
\label{NMHVsingleWL}
\mathcal{W}_{I,J}^{(1,0),\rm{c}} &= \sum_{i \in I} \sum_{j \in J} [*,i-1,i,j-1,j]\,, \notag \\
\mathcal{W}_{I,J}^{(2,0),\rm{c}} &= \frac{1}{2}\Bigl(\sum_{i \in I} \sum_{j \in J} [*,i-1,i,j-1,j] \Bigr)^2 = \frac{1}{2} \bigl(\mathcal{W}_{I,J}^{(1,0),\rm{c}}\bigr)^2\,.
\end{align}
In the above, e.g. \(i-1\) is understood modulo the ordered set \(I\). Here we give the contribution to the remainder function \(\mathcal{R}^{(2,1),\rm{c}}_{\{1,2,3,4\},\{5,6,7,8\}}\) which means that the coefficients of the zero-mass and two-mass easy boxes are not simply the leading singularities but rather have been modified by the subtraction of \(\mathcal{W}^{(2,0),\rm{c}}_{\{1,2,3,4\},\{5,6,7,8\}}(\mathcal{W}^{(0,1)}_{\{1,2,3,4\}}+\mathcal{W}^{(0,1)}_{\{5,6,7,8\}})\). Note that we choose a particular way to represent each box coefficient here, and other choices are certainly possible. 

For the contribution from zero-mass boxes, setting $P=\{(12),(23),(34),(41)\}$ we have 
\be
C_P \bar{\chi}_P + (\{1,2,3,4\} \leftrightarrow \{5,6,7,8\})\,.
\ee
where
\begin{align}
C_P &= \mathcal{W}^{(2,0),\rm{c}}_{\{1,2,3,4\},\{5,6,7,8\}} -\mathcal{W}^{(2,0),\rm{c}}_{\{1,2,3\},\{5,6,7,8\}} -\mathcal{W}^{(2,0),\rm{c}}_{\{2,3,4\},\{5,6,7,8\}}  -\mathcal{W}^{(2,0),\rm{c}}_{\{3,4,1\},\{5,6,7,8\}}  -\mathcal{W}^{(2,0),\rm{c}}_{\{4,1,2\},\{5,6,7,8\}} \notag \\
\bar{\chi}_{P} &= -3\Li_2(1) -\frac{1}{2}\biggl[\log\Bigl[\frac{\langle 12X \rangle \langle 34X\rangle}{\langle14X\rangle \langle 23X \rangle}\Bigr]\biggr]^2
\end{align}

There is no one-mass contribution for a correlator of two squares. For the two-mass easy contribution, setting $P=\{(12),(23),(56),(67)\}$ we have 
\be
C_{P} \bar{\chi}_P + \rm{cyc}_1 + \rm{cyc}_2\,,
\ee
where
\begin{align}
C_P = &F + 2\Bigl[\frac{\langle 2356\rangle \langle 6712 \rangle}{\langle2367\rangle\langle5612\rangle}-1\Bigr][(12) \cap (765), 2,3,5,6][(56)\cap(321),6,7,1,2] + \mathcal{W}_{\{1,2,3,4\},\{5,6,7,8\}}^{(2,0),\rm{c}}  \notag \\
\bar{\chi}_P = &\Li_2\Bigl[1-\frac{\langle 12X \rangle \langle 2367 \rangle}{\langle 23X \rangle \langle 1267 \rangle }\Bigr] + \Li_2\Bigl[1 - \frac{\langle 2356 \rangle \langle 1267 \rangle}{\langle 2367\rangle \langle 1256 \rangle }\Bigr] + \Li_2\Bigl[1-\frac{\langle 1256 \rangle \langle 67X\rangle}{\langle 56X \rangle \langle 1267\rangle}\Bigr]  \\
 - &\Li_2\Bigl[1 - \frac{\langle 12X\rangle \langle 2356\rangle}{\langle 23X\rangle \langle 1256 \rangle}\Bigr] - \Li_2\Bigl[1-\frac{\langle 2356 \rangle \langle 67X \rangle}{\langle 2367 \rangle \langle 56X \rangle}\Bigr] + \log\Bigl[\frac{\langle 12X \rangle \langle 2367 \rangle}{\langle 23X\langle 1267\rangle}\Bigr]\log\Bigl[\frac{\langle 1256 \rangle \langle 67X \rangle}{\langle 56X \rangle \langle 1267 \rangle}\Bigr]  \notag
\end{align}
where we define\footnote{This expression is really a limit of an N\(^2\)MHV single Wilson loop in a configuration where certain twistors appear more than once in the list of cusps; here we use a twistor diagrammatic expansion with a particular choice of reference twistor. Such terms will be introduced and explored in more detail in \cite{leadingSing}.}
{\small
\begin{align}
F=
-&[2,3,4,6,7][1,2,4,6,(67)\cap(234)]
-[2,3,4,7,8][1,2,4,7,(78)\cap(234)]\notag\\
-&[2,3,4,6,7][2,3,5,6,(34)\cap(267)]
-[2,3,4,7,8][2,3,5,6,(34)\cap(278)]\notag\\
+&[2,3,4,5,8][2,3,5,6,(34)\cap(285)]
+[2,3,4,6,7][2,3,5,8,(34)\cap(267)]\notag\\
+&[2,3,4,7,8][2,3,5,8,(34)\cap(278)]
-[2,3,4,6,7][2,3,7,8,(34)\cap(267)]\notag\\
-&[1,2,4,6,7][2,4,5,6,(41)\cap(267)]
-[1,2,4,7,8][2,4,5,6,(41)\cap(278)]\notag\\
+&[1,2,4,5,8][2,4,5,6,(41)\cap(285)]
+[1,2,4,6,7][2,4,5,8,(41)\cap(267)]\notag\\
+&[1,2,4,7,8][2,4,5,8,(41)\cap(278)]
-[1,2,4,6,7][2,4,7,8,(41)\cap(267)]\notag\\
+&[2,5,6,7,8][1,2,4,6,(67)\cap(285)]
-[2,5,6,7,8][1,2,4,7,(78)\cap(256)]\notag\\
+&[2,5,6,7,8][2,3,4,6,(67)\cap(285)]
-[2,5,6,7,8][2,3,4,7,(78)\cap(256)]\notag\\
-&[2,3,4,5,6][2,5,7,8,(56)\cap(234)]
-[1,2,4,5,6][2,5,7,8,(56)\cap(241)]\notag\\
+&[2,3,4,5,8][2,6,7,8,(85)\cap(234)]
+[1,2,4,5,8][2,6,7,8,(85)\cap(241)]\notag\\
+&[2,3,4,5,8][1,2,4,8,(85)\cap(234)]-[2,3,4,5,6][1,2,4,5,(56)\cap(234)]\notag\\
+&[1,2,4,5,8][2,3,4,5,6]
-[1,2,4,6,7][2,3,4,5,6]
-[1,2,4,7,8][2,3,4,5,6]\notag\\
+&[1,2,4,6,7][2,3,4,5,8]
+[1,2,4,7,8][2,3,4,5,8]
-[1,2,4,6,7][2,3,4,7,8]\notag\\
+&[1,2,4,5,6][2,5,6,7,8]
-[1,2,4,6,7][2,5,6,7,8]
+[2,3,4,5,6][2,5,6,7,8]\notag\\
-&[2,3,4,6,7][2,5,6,7,8]\notag
\end{align}
}
For the two-mass hard contribution, setting $P=\{(12_,(23),(34),(56)\}$ we have 
\be
C_P \bar{\chi}_P + (\{1,2,3,4\} \leftrightarrow \{5,6,7,8\}) + \rm{cyc}_1 + \rm{cyc}_2
\ee
where
\begin{align}
C_P&=[2,3,4,5,6]\Bigl[\mathcal{W}^{(1,0)}_{\{2,3,(43)\cap(562),(56)\cap(432),6,7,8,5,(56)\cap(432)\}} - \mathcal{W}^{(1,0)}_{\{(43)\cap(562),4,1,2,(56)\cap(432),6,7,8,5,(56)\cap(432)\}} \notag \\
&\,\,\, \quad \qquad \qquad +\mathcal{W}^{(1,0)}_{\{(12)\cap(653),2,3,(65)\cap(123),6,7,8,5,(65)\cap(1,2,3)\}}  - \mathcal{W}^{(1,0)}_{\{3,4,1,(12)\cap(653),(65)\cap(123),6,7,8,5,(65)\cap(123)\}} \notag \Bigr]\\
\bar{\chi}_P &= \Li_2\Bigl[1-\frac{\langle 12X\rangle \langle 2356\rangle}{\langle 1256 \rangle \langle 23X\rangle}\Bigr] - \Li_2\Bigl[1-\frac{\langle 23X\rangle \langle3456\rangle}{\langle 2356 \rangle \langle 34X\rangle}\Bigr] \notag \\
&- \frac{1}{2}\log \Bigl[\frac{\langle 12X\rangle\langle2356\rangle}{\langle1256\rangle\langle23X\rangle} \Bigr] \log \Bigl[\frac{\langle1256\rangle\langle34X\rangle}{\langle1234\rangle\langle56X\rangle}\Bigr]
+ \frac{1}{2}\log \Bigl[\frac{\langle 23X\rangle \langle 3456\rangle}{\langle 2356\rangle \langle 34X\rangle} \Bigr]\log \Bigl[\frac{\langle1234\rangle\langle56X\rangle}{\langle1256\rangle\langle34X\rangle} \Bigr] 
\end{align}

In the NMHV tree-level Wilson loops which enter into the above, note that certain twistors feature more than once. To interpret such terms, we can simply take (\ref{NMHVsingleWL}) literally, noting that any $R$-invariants with the same twistor featuring twice will vanish by antisymmetry. Note that this prescription actually gives something which is not reference twistor independent for the individual terms e.g. \(\mathcal{W}^{(1,0)}_{\{2,3,(43)\cap(562),(56)\cap(432),6,7,8,5,(56)\cap(432)\}}\) but multiplication by the $R$-invariant \([2,3,4,5,6]\) restores reference twistor independence. Strictly speaking, what is happening is that a decagonal Wilson loop is being sent to these configurations by a certain limit, under which certain terms in the diagrammatic expansion (\ref{NMHVsingleWL}) (which we are discarding in this prescription) actually \emph{diverge}. However, these divergent terms are eliminated for Grassmann reasons by the $R$-invariant \([2,3,4,5,6]\) and so in the overall expression they may safely be dropped. The provenance and proper definitions of such expressions will be explored in more detail in \cite{leadingSing}.

For the three-mass contributions, setting $P=\{(12),(23),(56),(78)\}$ we have
\be
C_P \bar{\chi}_P + \{1,2,3,4\} \leftrightarrow \{5,6,7,8\} + \rm{cyc}_1 + \rm{cyc_2}\,,
\ee
where
\begin{align}
C_P &= [2,5,6,7,8]\bigl(\mathcal{W}^{(1,0)}_{\{2,3,4,1,2,(65)\cap(872),6,7,8,(78)\cap(652)\}} - \mathcal{W}^{(1,0)}_{\{2,3,4,1,2,(78)\cap(872),8,5,(65)\cap(872)\}}\bigr) \notag \\
& + 2\Bigl[\frac{\langle 2356\rangle \langle 7812\rangle}{\langle 2378 \rangle \langle 5612 \rangle}-1\Bigr][(12) \cap (87 (65) \cap (321)), 2, 3, 5, 6][(65) \cap (321),7,8,1,2] \notag \\ 
\bar{\chi}_P &= \Li_2\Bigl[1-\frac{\langle 2356 \rangle \langle 1278 \rangle}{\langle 2378\rangle \langle 1256\rangle}\Bigr] - \Li_2\Bigl[1-\frac{\langle 12X \rangle \langle 2356 \rangle}{\langle 23X \rangle \langle 1256 \rangle}\Bigr] - \Li_2\Bigl[1 - \frac{\langle 23X \rangle \langle 1278\rangle}{\langle 12X \rangle \langle 2378 \rangle}\Bigr] \notag \\
&\frac{1}{2}\log \Bigl[\frac{\langle 12X \rangle \langle 2356 \rangle}{\langle 23X \rangle \langle 1256 \rangle} \Bigr] \log \Bigl[\frac{\langle 23X \rangle \langle 5678 \rangle}{\langle 2378\rangle \langle 56X\rangle}\Bigr] + \frac{1}{2}\log \Bigl[\frac{\langle 23X \rangle \langle 1278 \rangle}{\langle 12X \rangle \langle 2378 \rangle} \Bigr]\log \Bigl[\frac{\langle 12X\rangle\langle5678\rangle}{\langle 1256 \rangle \langle 78X \rangle}\Bigr]\,.
\end{align}

The same prescription as for the two-mass hard expressions applies to these NMHV single Wilson loops. Finally, for the four-mass contributions we have 
\be
C_P \bar{\chi}_P + \rm{cyc}_1 + \rm{cyc_2}\,,
\ee
where
\begin{align}
C_P &= x\phi_1 [\delta_1, 1,2,3,4][\beta_1,5,6,7,8] +\bar{x}\phi_2[\alpha_2,2,3,5,6][\gamma_2,7,8,1,2] \notag \\
\bar{\chi}_P &= \Li_2\bigl[\frac{1}{2}(1 - u + v + \Delta )\bigr]  +\Li_2\bigl[\frac{1}{2}(1 + u - v + \Delta )\bigr] - \Li_2(1)  \notag \\
& + \frac{1}{2}\log(u)\log(v) - \log\bigl[\frac{1}{2}(1 - u + v + \Delta )\bigr]\log\bigl[\frac{1}{2}(1 + u - v + \Delta)\bigr].
\end{align}
Here we define the cross ratios $u$ and $v$ and related variables $x$ and $\bar{x}$ via
\begin{align}
& u = x \bar{x} = \frac{\langle 1234\rangle \langle5678\rangle}{\langle1256\rangle\langle3478\rangle}\,,  &&v = (1-x)(1-\bar{x}) = \frac{\langle 3456\rangle \langle1278\rangle}{\langle1256\rangle\langle3478\rangle}\,, \notag \\
& x = \frac{1}{2}(\Delta - u + v -1)\,, &&\bar{x} = \frac{1}{2}( -\Delta - u + v - 1) \,,  \notag \\
&\Delta = x-\bar{x} = \sqrt{(1-u-v)^2-4uv}\,. && 
\end{align}
The super twistors $\alpha_2, \beta_1, \gamma_2, \delta_1$ are given by
\begin{align}
&\mathcal{Z}_{\alpha_2} = \mathcal{Z}_2 + \frac{\langle 234 (56)\cap(781)\rangle + \langle 134 (56) \cap (782)\rangle + \langle 1256 \rangle \langle 3478\rangle \Delta}{2\langle34 (56)\cap(781) 1\rangle}\mathcal{Z}_1\,, \notag \\
&\mathcal{Z}_{\gamma_2} = (56)\cap(78\alpha_2)\,, \notag \\
&\mathcal{Z}_{\beta_1} = \mathcal{Z}_3 + \frac{\langle 321 (78)\cap(654)\rangle + \langle 421 (78)\cap(653) \rangle + \langle 1256\rangle \langle3478 \rangle\Delta}{2\langle 21 (78) \cap (654) 4 \rangle}\mathcal{Z}_4\,, \notag \\
& \mathcal{Z}_{\delta_1} = (78)\cap(56\beta_1)
\end{align}
Finally we define the conformally invariant quantities $\phi_1$ and $\phi_2$,
\begin{align}
\phi_1 = \Bigl(1-\frac{\langle \beta_1 8 12 \rangle \langle \delta_1 4 56\rangle}{\langle \beta_1 856 \rangle \langle \delta_1 4 12 \rangle}\Bigr)^{-1} \qquad \phi_2 =  \Bigl(1-\frac{\langle \alpha_2 678 \rangle \langle \gamma_2 234  \rangle}{\langle \alpha_2 6 34 \rangle \langle \gamma_2 2 78 \rangle} \Bigr)^{-1}
\end{align}

Summing over all the above contributions yields \(\mathcal{R}^{(2,1)}_{\{1,2,3,4\},\{5,6,7,8\}}\). Note that the independence on the auxillary twistor \(X\) is not immediately manifest but, upon computing derivatives, can be seen by making use of a number of simple linear relations\footnote{Among the \(70\) box coefficients, there are \(33\) simple linear relations.} which exist among the various leading singularities. 

\section{\(\bar{Q}\)-equation beyond the planar limit}
\label{Sec-nonPlanarQbar}
Since the connected part of the correlation function of two Wilson loops is colour-suppressed relative to the connected parts, any test of the generalised \(\bar{Q}\)-equation for multiple Wilson loops naturally requires one to go beyond the \(N \to \infty\) limit. The original \(\bar{Q}\)-equation for a single Wilson loop (here we pluck out the term at fixed MHV degree) reads
\be
\bar{Q}_A^{A'} \mathcal{R}^{(k)}_n = \frac{1}{4}\Gamma_{\rm cusp} \int [d^{2|3} \mathcal{Z}_{n+1}]^{A'}_A \bigl[ \mathcal{R}^{(k+1)}_{n+1} - \mathcal{R}^{(k)}_n \mathcal{R}^{(1,0)}_{n+1}\bigr] + \text{ cyc}\,.
\label{Qbar}
\ee
Recall that in terms of momentum supertwistors,
\be
\bar{Q}^{A'}_A = \sum_{i} \chi_i^{A'}\frac{\partial}{\partial Z_{i}^{A}}.
\ee
The integration measure in (\ref{Qbar}) is defined by
\be 
\int \bigl[d^{2|3}\mathcal{Z}_{{n}+1}\bigr]^{A'}_A = C(n-1 n 1)_A  \oint_{\epsilon=0} \epsilon d \epsilon \int_0^{\infty} d\tau (d^3 \chi_{n+1})^{A'} 
\ee
where we send \(\mathcal{Z}_{n+1}\) collinear with \(\mathcal{Z}_n\) via
\be
\mathcal{Z}_{n+1} = \mathcal{Z}_n - \epsilon \mathcal{Z}_{n-1} + C \epsilon \tau \mathcal{Z}_1 + C' \epsilon^2 \mathcal{Z}_2
\ee
with
\be
C = \frac{\langle n-1 \, n \, 2 \, 3}{\langle n123\rangle} \, \quad \quad  C' = \frac{\langle n-2 \, n-1 \, n \, 1}{\langle n-2 \, n-1 \, 2 \, 1\rangle}.
\ee
and where \(\epsilon \to 0\) parametrises the collinear limit and \(\tau\) is the longitudinal momentum fraction to be integrated over. This equation was originally argued for in \cite{Caron-Huot:2011dec} using Wilson loop arguments. As such, one would expect that, beyond the planar limit for which the amplitude and Wilson loop coincide, the equation should hold for the colour-exact Wilson loop, not the colour-exact amplitude. Since tests of the \(\bar{Q}\)-equation in the literature have primarily been interested in the dual amplitudes and have as such restricted to the planar limit, let us convince ourselves that the equation indeed holds for colour-exact Wilson loops computed in our conventions.

At leading order in \(g^2\), we have
\[\frac{1}{4}\Gamma_{\rm cusp} = \frac{2C_F}{N}g^2\]
which of course reduces to simply \(g^2\) in the planar limit. Inspecting 
(\ref{Qbar}) at $O(g^2)$, we thus see that the equation relates the action of \(\bar{Q}^{A'}_A\) on \(\mathcal{R}_n^{(k+1,1)}\) at one-loop to the collinear integral over tree-level N\(^k\)MHV Wilson loops on the right-hand side. 

\subsection{An explicit check for \(k=1\)}

Let us restrict our attention to the case \(k=1\), for which the \(\bar{Q}\)-equation at $O(g^2)$ reads 
\be
\bar{Q}_A^{A'} \mathcal{R}^{(1,1)}_{n+1} = \frac{2C_F}{N} \int \bigl[d^{2|3} \mathcal{Z}_{n+1}\bigr]^{A'}_A \bigl[ \mathcal{R}^{(2,0)}_{n+1} - \mathcal{R}^{(1,0)}_n \mathcal{R}^{(1,0)}_{n+1}\bigr] + \text{ cyc}\,.
\label{Qbarspecific}
\ee
On the right-hand side, \(\mathcal{R}_n^{(1,0)}\) and \(\mathcal{R}_{n+1}^{(1,0)}\) are the same in the planar and non-planar theories (as there is only a single propagator per diagram), and each diagram comes with a global factor of \(1\). The other term, \(\mathcal{R}_{n+1}^{(2,0)}\), receives contribution from planar diagrams where propagators do not cross, and non-planar diagrams where propagators do cross. The planar diagrams come with factor (this includes contributions from the normalisation of the loop operator and propagators as well as the colour trace)
\be
c_p = 1\,,
\ee
while the non-planar diagrams come with factor
\be
c_n = -\frac{1}{2NC_F} = - \frac{1}{N^2-1}\,.
\ee

We may conveniently express the full, non-planar \(\mathcal{R}_{n+1}^{(2,0)}\) in terms of the planar and Abelian contributions, and thus invoke the planar and Abelian \(\bar{Q}\)-equations (which are known to hold) to express the right-hand side in terms of the action of \(\bar{Q}\) instead of a collinear integral. Let us write \(\mathcal{D}_p\) to denote the sum of the colour-stripped\footnote{Since in our normalisation the colour factor is in fact \(1\), this can equivalently be thought of as simply the planar observable.} planar diagrams contributing to \(\mathcal{R}_{n+1}^{(2,0)}\), and \(\mathcal{D}_n\) to mean the sum of the colour-stripped non-planar diagrams. In this notation, the colour-exact expression for \(\mathcal{R}_{n+1}^{(2,0)}\) is 
\be
\mathcal{R}_{n+1}^{(2,0)} = c_p \mathcal{D}_p + c_n \mathcal{D}_n.
\ee
If we further define \(\mathcal{D}_a\) as the sum of all Abelian diagrams (all of which come with unit factor), we can write
\be
\mathcal{D}_n = \mathcal{D}_a - \mathcal{D}_p.
\ee
Making this replacement we have
\begin{align}
\mathcal{R}_{n+1}^{(2,0)} = & (c_p-c_n) \mathcal{D}_p + c_n \mathcal{D}_a \notag \\
 = & \frac{N^2}{N^2-1} \mathcal{D}_p  - \frac{1}{N^2-1} \mathcal{D}_a 
\end{align}
The other term on the right-hand side, 
\(\mathcal{R}_n^{(1,0)}\mathcal{R}_{n+1}^{(1,0)}\), is the same in $SU(N)$ (whether \(N \to \infty\) or not) and the Abelian theory. The expression to be subjected to the collinear integration on the right-hand side of (\ref{Qbarspecific}) is 
\be
\frac{N^2}{N^2-1}\mathcal{D}_p - \frac{1}{N^2-1} \mathcal{D}_a - \mathcal{R}_n^{(1,0)}\mathcal{R}_{n+1}^{(1,0)}
\ee
which we may helpfully rewrite as 
\be
\frac{N^2}{N^2-1}(\mathcal{D}_p - \mathcal{R}_n^{(1,0)}\mathcal{R}_{n+1}^{(1,0))}) -\frac{1}{N^2-1} (\mathcal{D}_a - \mathcal{R}_n^{(1,0)}\mathcal{R}_{n+1}^{(1,0)}).
\ee
Now, note that we can use the planar and Abelian \(\bar{Q}\) equations to rewrite (note that in the planar and Abelian \(\bar{Q}\) equations, \(\frac{1}{4}\Gamma_{\rm cusp}\) reduces to \(g^2\))
\be
\int \bigl[d^{2|3}\mathcal{Z}_{n+1}\bigr]^{A'}_A[\mathcal{D}_p - \mathcal{R}_n^{(1,0)}\mathcal{R}_{n+1}^{(1,0)}] = \bar{Q}_{A}^{A'} \mathcal{R}_{n+1}^{(1,1), \textrm{planar}}
\ee
and
\be
\int \bigl[d^{2|3}\mathcal{Z}_{n+1}\bigr]^{A'}_A[\mathcal{D}_a - \mathcal{R}_n^{(1,0)}\mathcal{R}_{n+1}^{(1,0)}] = \bar{Q}_{A}^{A'} \mathcal{R}_{n+1}^{(1,1), \textrm{abelian}} = 0.
\ee
Upon making use of this identification, we can see that (\ref{Qbarspecific}) simply amounts to the claim that 
\be
\bar{Q}_{A}^{A'} \mathcal{R}_{n+1}^{(1,1)} = \frac{2C_F}{N} \frac{N^2}{N^2-1}\bar{Q}_{A}^{A'}\mathcal{R}_{n+1}^{(1,1), \textrm{planar}} = \bar{Q}_{A}^{A'}\mathcal{R}_{n+1}^{(1,1), \textrm{planar}}
\ee
where on the right-hand side the \(\frac{2C_F}{N}\) comes from the colour-exact expression for \(\Gamma_{\rm cusp}\) and the \(\frac{N^2}{N^2-1}\)
comes from the coefficient of the term which we rewrote using the planar \(\bar{Q}\) equation. 

Therefore we need only verify that
\be
\bar{Q}_A^{A'} \mathcal{R}_{n+1}^{(1,1)} = \bar{Q}_A^{A'}\mathcal{R}_{n+1}^{(1,1), \textrm{planar}}\,.
\ee
In fact, this holds because
\be
\mathcal{R}_{n+1}^{(1,1)} = \mathcal{R}_{n+1}^{(1,1), \textrm{planar}}
\ee
which is itself equivalent to (here we use that all products of diagrams in the expansion of \(\mathcal{W}^{(1,0)}_{n+1}\mathcal{W}^{(0,1)}_{n+1}\) come with a universal colour factor of \(1-\frac{1}{N^2} = \frac{N^2-1}{N^2}\), which reduces to \(1\) in the planar limit)
\be
\mathcal{W}_{n+1}^{(1,1)} - \mathcal{W}^{(1,0)}_{n+1}\mathcal{W}^{(0,1)}_{n+1}= \mathcal{W}_{n+1}^{(1,1), \textrm{planar}} - \frac{N^2}{N^2-1}\mathcal{W}^{(1,0)}_{n+1}\mathcal{W}^{(0,1)}_{n+1}
\ee
Note that \(\mathcal{W}^{(1,0)}_{n+1}\mathcal{W}^{(0,1)}_{n+1}\) is the colour-exact expression which we convert to the planar limit on the right-hand side by dividing by its prefactor \(\frac{N^2-1}{N^2}\). This simplifies to 
\be
\mathcal{W}_{n+1}^{(1,1)} + \frac{1}{N^2-1}\mathcal{W}^{(1,0)}_{n+1}\mathcal{W}^{(0,1)}_{n+1}= \mathcal{W}_{n+1}^{(1,1), \textrm{planar}} 
\label{relation}
\ee
To see that this holds, note that at the level of twistor diagrams the first colour-exact object on the left receives contributions from diagrams with two insertions on the Lagrangian, and diagrams with three insertions on the Lagrangian. Diagrams with three insertions on the Lagrangian (which means each propagator runs from the Lagrangian line to a distinct twistor line in the Wilson loop)
can come with planar factor \(\frac{N^2 - 2}{N^2}\) or non-planar factor \(-\frac{2}{N^2}\). Each non-planar diagram is related to a planar diagram by a non-cyclic permutation of the order of the propagator insertions on the Lagrangian, which - since there are only three - is equivalent to \emph{reflecting} the order of the insertions. As was discussed in \cite{Drummond:2025ulh} in the context of the simplification in the Abelian theory, such a reflection with an odd number of insertions on the Lagrangian means the two diagrams differ by a factor of \(-1\), and so we can combine each non-planar diagram with its planar partner. Since
\be
\frac{N^2-2}{N^2} - \Bigl(-\frac{2}{N^2}\Bigr) = 1,
\ee
we can see that the contributions with a three-vertex are equivalent to taking only such planar diagrams and dressing them with a factor of \(1\) (as is indeed the factor the planar diagrams are dressed with in the large \(N\) limit). 

Similarly, diagrams with two insertions on the Lagrangian can come with planar factor $(1-\frac{1}{N^2})$ or non-planar factor \(-\frac{1}{N^2}\). When we add the term \(\frac{1}{N^2-1}\mathcal{W}_{n+1}^{(1,0)}\mathcal{W}_{n+1}^{(0,1)}\), there is the addition of an expression kinematically identical (by which we mean identical up to the \(N\)-dependent pre-factor) to each of these diagrams\footnote{In the case of diagrams without a double insertion on a twistor line other than the Lagrangian, there is a single product of diagrams in \(\mathcal{W}_{n+1}^{(1,0)}\mathcal{W}_{n+1}^{(0,1)}\) which is kinematically equivalent; for those with a double insertion, one product of diagrams in \(\mathcal{W}_{n+1}^{(1,0)}\mathcal{W}_{n+1}^{(0,1)}\) is kinematically equivalent to the sum of the two versions of a diagram with a double insertion on a twistor line.}.

Since (here \(\frac{1}{N^2-1}\) multiplies that term on the right-hand side of (\ref{relation}) and $(1-\frac{1}{N^2})$ is the factor each product of diagrams in  \(\mathcal{W}_{n+1}^{(1,0)}\mathcal{W}_{n+1}^{(0,1)}\) comes with)
\be
\frac{1}{N^2-1} \times \Bigl(1 - \frac{1}{N^2}\Bigr) = \frac{1}{N^2}
\ee
this eliminates the non-planar diagrams and converts the overall factor of the planar diagrams to \(1\), as required for a match with the right-hand side of (\ref{relation}).

We thus see that, with our choice of normalisation, the \(\bar{Q}\)-equation holds for \(k=1\) and all \(n\), for the colour-exact Wilson loop up to order $g^2$. Note that the \(\bar{Q}\)-equation is certainly sensitive to our choice of scaling of \(S_1\) and \(S_2\) as shown explicitly in (\ref{selfdual}) and (\ref{interaction}). Such a scaling effectively rescales the Grassmann variables and hence an extra factor would need to be included in the $\bar{Q}$-equation. With our choice of action the equation (\ref{Qbarspecific}) holds as written.

Computing the right-hand side of the \(\bar{Q}\)-equation for \(k>1\) becomes more challenging. However, the non-planar $\bar{Q}$ equation (\ref{Qbar}) can still be verified at higher MHV degree by making use of a colour-exact Wilson loop recursion relation in the $SU(N)$ theory which is obtained by closely following the procedure of \cite{Bullimore:2011ni} (and which we will discuss in more detail in forthcoming work \cite{multiBCFW}). With this recursion relation, the process amounts to adapting the procedure spelled out in \cite{Caron-Huot:2011dec} for integrating the RHS of the planar $\bar{Q}$ equation. Details are included in Appendix \ref{rhsqbarnonplanar} and we used this procedure to successfully perform another check for \(n=7\), \(k=2\).

\section{\(\bar{Q}\)-equation for multiple Wilson loops}

In \cite{Drummond:2025ulh}, we conjectured a generalisation of the \(\bar{Q}\)-equation for correlators of multiple light-like Wilson loops, and the validity of this equation in the simple case of the Abelian theory was checked. The conjectured equation is 
\begin{equation}
    \bar{Q}^{A'}_{A} \mathcal{R}_{n_1,\ldots, n_m}= \frac{1}{4}\Gamma_{\textrm{cusp}} \sum_{r=1}^m\left[\int \left[d^{2|3}\mathcal{Z}_{X_r}\right]^{A'}_{A}\left[\mathcal{R}_{n_1,\ldots,n_r+1,\ldots,n_m}-\mathcal{R}_{n_1,\ldots,n_m}\mathcal{R}_{n_r+1}^{(1,0)}\right]+\textrm{cyc}_r\right]
    \label{QbarMWL}
\end{equation}
where $\{n_1,\ldots,n_m\}$ are the numbers of sides on each of $m$ Wilson loops, where the twistors on Wilson loop $r$ are labelled by $\{1^r,\hdots,n_r^r\}$. The superscript denotes the Wilson loop which the given twistor is on. For the twistors below which have this index suppressed, we are referring to the twistors on the $r$-th Wilson loop (unless stated otherwise).

On the right-hand side, an extra vertex \(\mathcal{Z}_{X_r}\) is added after \(\mathcal{Z}_{n_r}\) on the \(r\)'th Wilson loop, which is then subjected to the collinear limit equivalent to the one used for a single Wilson loop, 
\be
\mathcal{Z}_{X_r} \to \mathcal{Z}^r_{n_r} - \epsilon \mathcal{Z}^r_{n_r-1} + C \epsilon \tau \mathcal{Z}^r_{1} + C' \epsilon^2  \mathcal{Z}^r_{2},
\ee
where again \(\epsilon \to 0\) gives the collinear limit and \(\tau\) is the longitudinal momentum fraction to be integrated over, and where we have the bosonic factors
\be
C=\frac{\langle n_r -1 , n_r , 2^r, 3^r \rangle }{\langle n_r , 1^r , 2^r , 3^r\rangle }\,, \qquad
C'=\frac{\langle n_r -2 , n_r -1, n_r, 1^r \rangle }{\langle  n_r -2 , n_r-1 , 2^r , 1^r\rangle }\,.
\ee
The `\(+\textrm{cyc}_r\)' then amounts to summing over all the possible positions for the insertion of the extra vertex on this Wilson loop. Finally, the integration measure is given explicitly as 
\be 
\int \bigl[d^{2|3}\mathcal{Z}_{X_r}\bigr]^{A'}_A = C(n_r-1, n_r , 1^r)_A  \oint_{\epsilon=0} \epsilon d \epsilon \int_0^{\infty} d\tau (d^3 \chi_{X_r})^{A'}.
\ee

Although the equation (\ref{QbarMWL}) is expected to hold non-perturbatively, it is convenient to inspect the contribution to each side at a given order in \(g^2\) and given MHV degree. In the following we may occassionally suppress the residue operation and the limits of the $\tau$ integral.

Let us now perform an explicit check of the \(\bar{Q}\)-equation in the case of two Wilson loops in the $SU(N)$ theory. The \(\bar{Q}\) equation for two Wilson loops reads 
\begin{align}
    \bar{Q}^{A'}_{A} \mathcal{R}_{n_1, n_2}= \frac{1}{4}\Gamma_{\textrm{cusp}}\biggl[&\int \left[d^{2|3}\mathcal{Z}_{X_1}\right]^{A'}_{A}\left[\mathcal{R}_{n_1+1,n_2}-\mathcal{R}_{n_1,n_2}\mathcal{R}_{n_1+1}^{(1,0)}\right]+\textrm{cyc}_1 \notag \\
    +&\int \left[d^{2|3}\mathcal{Z}_{X_2}\right]^{A'}_{A}\left[\mathcal{R}_{n_1,n_2+2}-\mathcal{R}_{n_1,n_2}\mathcal{R}_{n_2+1}^{(1,0)}\right]+\textrm{cyc}_2  \biggr]
\end{align}
We recall that $\mathcal{R}_{n_1,n_2}$ and $\frac{1}{4}\Gamma_{\textrm{cusp}}$ have the following small coupling expansions,
\begin{equation}
    \frac{1}{4}\Gamma_{\textrm{cusp}}=\frac{2C_F}{N}g^2+O(g^4) \hspace{10mm} \mathcal{R}_{n_1,n_2}^{(k)}=\mathcal{R}^{(k,0)}_{m,n}+g^2\mathcal{R}^{(k,1)}_{n_1,n_2}+g^4\mathcal{R}^{(k,2)}_{n_1,n_2}+O(g^6)
\end{equation}
where we recall $g^2=\frac{g_{YM}^2N}{16\pi^2}$. Inspecting each side of the equation at $O(g^2)$ and N\(^k\)MHV, we obtain a relation which should hold between the N\(^{k+1}\)MHV tree result and the (integrated) N\(^k\)MHV $O(g^2)$ result,
\begin{align}
    \bar{Q}^{A'}_{A} \mathcal{R}^{(k,1)}_{n_1, n_2}=\quad &\int \left[d^{2|3}\mathcal{Z}_{X_1}\right]^{A'}_{A}\left[\mathcal{R}_{n_1+1,n_2}^{(k+1,0)}-\mathcal{R}_{n_1,n_2}^{(k,0)}\mathcal{R}_{n_1+1}^{(1,0)}\right]+\textrm{cyc}_1\notag \\
    +&\int \left[d^{2|3}\mathcal{Z}_{X_2}\right]^{A'}_{A}\left[\mathcal{R}_{n_1,n_2+1}^{(k+1,0)}-\mathcal{R}_{n_1,n_2}^{(k,0)}\mathcal{R}_{n_2+1}^{(1,0)}\right]+\textrm{cyc}_2 \,.
\end{align}

Recall that the leading order contribution to the correlator of two Wilson loops in \(N\) is the disconnected part. Since we have satisfied ourselves that the \(\bar{Q}\)-equation does hold for a single Wilson loop even when retaining the full colour structure, we can see that the above equation is equivalent to the same equation isolated to the connected part only. Denoting the connected part with a superscript 'c' for notational brevity, this means we need to verify that 
\begin{align}
    \bar{Q}^{A'}_{A} \mathcal{R}^{(k,1),{\rm c}}_{n_1, n_2}=\quad  &\int \left[d^{2|3}\mathcal{Z}_{X_1}\right]^{A'}_{A}\left[\mathcal{R}_{n_1+1,n_2}^{(k+1,0),{\rm c}}-\mathcal{R}_{n_1,n_2}^{(k,0), {\rm c}}\mathcal{R}_{n_1+1}^{(1,0)}\right]+\textrm{cyc}_1 \notag \\
    +&\int \left[d^{2|3}\mathcal{Z}_{X_2}\right]^{A'}_{A}\left[\mathcal{R}_{n_1,n_2+1}^{(k+1,0),{\rm c}}-\mathcal{R}_{n_1,n_2}^{(k,0),{\rm c}}\mathcal{R}_{n_2+1}^{(1,0)}\right]+\textrm{cyc}_2\,.
    \label{QbarMWLconn}
\end{align}

It is relation (\ref{QbarMWLconn}) which we now seek to check in some simple cases by explicit computation. We specialise here to the large \(N\) limit, i.e. considering the leading order contributions in \(N\) to the connected part. To make the notation more explicit, we label the Wilson loop correlators on the right-hand side by their twistor vertices. Instead of using a superscript to differentiate between twistors on different loops, we label the vertices of the first Wilson loop by $\{1,\ldots,n_1\}$ and the vertices of the second Wilson loop by $\{\tilde{1},\ldots,\tilde{n}_2\}$. Equation (\ref{QbarMWLconn}) then becomes
\begin{alignat}{2}
        & \bar{Q}_A^{A'}\mathcal{R}_{n_1,n_2}^{(k,1), {\rm c}}\nonumber\\
        &= \int \bigl[d^{2|3}Z_{X_1}\bigr]_A^{A'}\big[\langle\mathcal{L}[1,\ldots,n_1,X_1]\mathcal{L}[\tilde{1},\ldots,\tilde{n}_2]\rangle^{(k+1,0),{\rm c}}\nonumber\\
        &\hspace{31mm}-\langle\mathcal{L}[1,\ldots,n_1]\mathcal{L}[\tilde{1},\ldots,\tilde{n}_2]\rangle^{(k,0),{\rm c}}\langle\mathcal{L}[1,\ldots,n_1,X_1]\rangle^{(1,0)}\big]+\textrm{cyc}_1\nonumber\\
        &\hspace{4mm}+\int \bigl[d^{2|3}Z_{X_2}\bigr]_A^{A'}\big[\langle\mathcal{L}[1,\ldots,n_1]\mathcal{L}[\tilde{1},\ldots,\tilde{n}_2,X_2]\rangle^{(k+1,0),{\rm c}}\nonumber\\
        &\hspace{35mm}-\langle\mathcal{L}[1,\ldots,n_1]\mathcal{L}[\tilde{1},\ldots,\tilde{n}_2]\rangle^{(k,0), {\rm c}}\langle\mathcal{L}[\tilde{1},\ldots,\tilde{n}_2,X_2]\rangle^{(1,0)}\big]+\textrm{cyc}_2\,.
        \label{qbar1loop}
\end{alignat}
In (\ref{qbar1loop}) $Z_{X_1}$ in the first term is parametrised as follows,
\begin{equation}
    Z_{X_1}=Z_{n_1}-\epsilon Z_{n_1-1}+C \epsilon \tau Z_1+C' \epsilon^2Z_2
\end{equation}
where $C=\frac{\langle n_1-1\hspace{1mm} n_1 \hspace{1mm}2 \hspace{1mm} 3 \rangle}{\langle n_1\hspace{1mm} 1\hspace{1mm} 2 \hspace{1mm}3\rangle}$ and $C'=\frac{\langle n_1-2 \hspace{1mm}n_1-1\hspace{1mm} n_1\hspace{1mm} 1 \rangle}{\langle n_1-2\hspace{1mm} n_1-1\hspace{1mm} 2\hspace{1mm} 1 \rangle}$. The collinear limit is $\epsilon\rightarrow 0$.  $Z_{X_2}$ is parameterised in an analogous way for the terms with $Z_{X_2}$ placed in the different positions along the second Wilson loop. 
To verify this equation, we use the chiral box expansion form of $\mathcal{R}_{n_1,n_2}^{k,1\textrm{-loop}}$ and use a BCFW-like recursion relation for connected Wilson loop correlators, to appear in \cite{multiBCFW}, to expand the correlators on the RHS of (\ref{qbar1loop}). Below, we outline the procedure for computing the LHS and RHS of (\ref{qbar1loop}).

\subsection{Computing the left-hand side}
\label{lhsqbar}

To compute the LHS of (\ref{qbar1loop}), we use the chiral box expansion of the 1-loop remainder function, 
\begin{equation}
\mathcal{R}_{n_1,n_2}^{(k,1),{\rm c}} = \sum_{P} C_P^k \bar{\chi}_{P},
\end{equation}
writing \(\bar{\chi}_{P}\) for the integrated chiral box with physical poles \(P\), and \(C_P^k\) for the coefficient, which will be the difference of the cuts of the integrand (in the N\(^k\)MHV sector) on the two Schubert solutions (as at the level of the integrand, one chirality is dressed with plus the cut on one Schubert solution, and the other is dressed with minus the cut on the other Schubert solution), modified by the addition of the \(N^k\)MHV tree result for zero mass, one mass and two-mass easy boxes which stay in the same Wilson loop due to the conversion to the remainder function.

Note that of course that the chiral box integrals \(\bar{\chi}_P\) are the same regardless of MHV degree, while the coefficients will differ (being straightforward to compute by taking residues, or alternatively using the formulae in terms of tree-level objects to appear in companion paper \cite{leadingSing}). All of the coefficients are annihilated by \(\bar{Q}\), meaning \(\bar{Q}\) passes through the coefficients and acts on the chiral box integrals only. We thus have 
\begin{equation}
\bar{Q}_A^{A'} \mathcal{R}_{n_1,n_2}^{(k,1),{\rm c}} = \sum_{P} C^k_{P} \bar{Q}^{A'}_A \bar{\chi}_P\,.
\end{equation}

The action of \(\bar{Q}\) on the chiral box integrals is straightforward owing to the fact that these are formed entirely from dilogarithms and products of logarithms. Recalling that the total derivative of the dilogarithm can be written as $d \,{\rm Li}_2(x) = -\log(1-x) d \log x$, we have that
\begin{equation}
\bar{Q}_A^{A'} \textrm{Li}_{2}(f) = -\log(1-f) \bar{Q}_A^{A'}\log(f)
\label{qbaract1}
\end{equation}
and
\begin{equation}
\bar{Q}_A^{A'} \log(f)\log(g) = \log(f)\bar{Q}_A^{A'}\log(g) + \log(g)\bar{Q}_A^{A'}\log(f). 
\label{qbaract2}
\end{equation}
In this way the left-hand side of the \(\bar{Q}\) equation may be straightforwardly expressed as a linear combination of terms of the form \(\log(f)\bar{Q}_A^{A'}\log(g)\), dressed with coefficients which are simply the coefficients \(C_P\) in the chiral box expansion for the remainder function. 

\subsection{Computing the right-hand side}
\label{rhsqbarplanar}

We have two objects to integrate on the RHS of (\ref{qbar1loop}), $\langle \mathcal{L}[1,\ldots,n_1,X_1]\mathcal{L}[\tilde{1},\ldots,\tilde{n}_2]\rangle^{(k+1,0),{\rm c}}$ and $\langle\mathcal{L}[1,\ldots,n_1,X_1]\rangle^{(1,0)}$, and the equivalent terms coming from inserting \(X_2\) on the second Wilson loop; the collinear integration passes through correlators which don't contain \(X_1\) or \(X_2\). 

The term $\langle\mathcal{L}[1,\ldots,n_1,X_1]\rangle^{(1,0)}$ can be expressed as a sum over $R$-invariants, each of which has dependence on $X_1$, where none of the twistors appearing are shifted.
\begin{equation}
        \langle\mathcal{L}[1,\ldots,n_1,X_1]\rangle^{(1,0)}=\sum_{i=1}^{n_1-1}\sum_{\substack{j=i+2,\\j \neq i-1,i,i+1}}^{X_1}[*,i-1,i,j-1,j]
        \label{singleWL}
\end{equation}
Let us choose $*=n-1$. For a term in this expression to contribute to the collinear integral, $n$ and $X_1$ must both appear in the $R$-invariant (otherwise, it would have no $\epsilon$ poles). This occurs only for $j=X_1$. Thus, this factor in the $\bar{Q}$ equation becomes
\begin{equation}
 \sum_{i=2}^{n-2}[n-1,i-1,i,n,X_1]
 \label{contribterm}
\end{equation}
The collinear integration acts simply on these terms and indeed the result for each type of $R$-invariant which can appear is given in \cite{Caron-Huot:2011dec}. After collinear integration these terms thus reduce to expressions of the form \(\log(f)\bar{Q}_A^{A'}\log(g)\) dressed with the tree-level expressions which had multiplied them in the right-hand side of (\ref{qbar1loop}). The collinear integral of the correlator requires some more work. Following the analagous procedure for a single Wilson loop spelled out in \cite{Caron-Huot:2011dec}, we use a particular BCFW shift to expand it in terms of other correlators, each of which has an overall $R$-invariant factor. Performing the shift $Z_{n_1}\rightarrow Z_{n_1}+tZ_{n_1-1}$ gives an expression in which the correlators can be sent to their collinear limit and only the single $R$-invariant factors out front need to be integrated. The expression is given by\footnote{While we defer a more detailed exploration of this formula to forthcoming work, let us briefly comment on the term $\langle\mathcal{L}[X_1,1,\ldots,\hat{n}_{1,\tilde{\jmath}},I_{\tilde{\jmath}},\tilde{\jmath},\ldots,\tilde{\jmath}-1,I_{\tilde{\jmath}}]\rangle$, which is a Wilson loop in a self-intersecting configuration. This can be defined by first taking the Wilson loop with the appropriate number of cusps, first identifying the second instance of \(\mathcal{Z}_{I_{\tilde{j}}}\) with \(\mathcal{Z}_{I_{\tilde{j}}} + \eta \mathcal{Z}_r\) for an arbitrary super-twistor \(\mathcal{Z}_r\), and then taking the limit \(\eta \to 0\) supersymmetrically. In fact, the object diverges under such a limit (even at tree level), but if one multiplies by the $R$-invariant pre-factor \emph{before} taking the limit, the divergences are eliminated for Grassmann reasons. The resulting finite limit can be verified directly by comparing to the tree-level expression obtained from diagrams.}
\begin{align}
    \langle \mathcal{L}[1,\ldots,n_1,X_1]\mathcal{L}[\tilde{1},\ldots,\tilde{n}_2]\rangle^{(k+1,0),{\rm c}}=&\langle \mathcal{L}[1,\ldots,n_1]\mathcal{L}[\tilde{1},\ldots,\tilde{n}_2]\rangle^{(k+1,0),{\rm c}} \notag \\
    &+\sum_{i=2}^{n_1-2}[n_1-1,n_1,X_1,i-1,i] \, T_i \notag \\
    &+\sum_{\tilde{\imath}=\tilde{1}}^{\tilde{n}_2} \,\, [n_1-1,n_1,X_1,\tilde{\imath}-1,\tilde{\imath}]\, \tilde{T}_i\,.
    \label{BCFW}
\end{align}
Here we have defined
\begin{align}
T_i &= \sum_{k'=0}^{k} \Big[\langle\mathcal{L}[X_1,1,\ldots,i-1,I_i]\mathcal{L}[\tilde{1},\ldots,\tilde{n}_2]\rangle^{(k',0),{\rm c}}\langle\mathcal{L}[I_i,i,\ldots,n_1-1,\hat{n}_{1,i}]\rangle^{(k-k',0)} \notag \\
    &\qquad  +\langle\mathcal{L}[X_1,1,\ldots,i-1,I_i]\rangle^{(k',0)}\langle \mathcal{L}[\tilde{1},\ldots,\tilde{n}_2] \mathcal{L}[I_i,i,\ldots,n_1-1,\hat{n}_{1,i}]\rangle^{(k-k',0),{\rm c}}\Big] \,,\\
\tilde{T}_i &= \langle\mathcal{L}[X_1,1,\ldots,\hat{n}_{1,\tilde{\imath}},I_{\tilde{\imath}},\tilde{\imath},\ldots,\tilde{\imath}-1,I_{\tilde{\imath}}]\rangle^{(k,0)} \notag \\
&\qquad -\alpha \sum_{k'=0}^{k}\langle\mathcal{L}[X_1,1,\ldots,\hat{n}_{1,\tilde{\imath}}]\rangle^{(k',0)}\langle \mathcal{L}[\tilde{1},\ldots,\tilde{n}_2]\rangle^{(k-k',0)}
\end{align}
and
\begin{align}
    &I_i=(i-1\hspace{1mm}i)\cap(n_1-1\hspace{1mm}n_1\hspace{1mm}X_1)\,, \hspace{3mm} && Z_{\hat{n}_{1,i}}=(n_1-1\hspace{1mm}n_1)\cap (i-1\hspace{1mm}i\hspace{1mm}X_1)\,, \hspace{3mm}  \notag \\
    &I_{\tilde{\imath}}=(\tilde{\imath}-1\hspace{1mm}\tilde{\imath})\cap(n_1-1\hspace{1mm}n_1\hspace{1mm}X_1)\,,  &&Z_{\hat{n}_{1,\tilde{\imath}}}=(n_1-1\hspace{1mm}n_1)\cap (\tilde{\imath}-1\hspace{1mm}\tilde{\imath}\hspace{1mm}X_1)\,.
    \label{intersectpoints}
\end{align}
For our present case of $SU(N)$, $\alpha = 1$. For $U(N)$, \(\alpha =0\). The collinear integration only acts on the $R$-invariant factors and the correlator factors can simply be set to their collinear limit, which amounts to sending their twistors to their collinear limit. This can be understood by considering the different cases appearing in (\ref{BCFW}).

Case 1: The Wilson loops in (\ref{BCFW}) with dependence on $X_1$ but not on $\hat{n}_{1,\tilde{\imath}}$ are $O(\epsilon^0)$ and those terms do not contribute poles if any of the $\chi_{X_1}$ integration acts on the Wilson loop factor. The twistors are simply sent to their collinear limit and there will be no $\tau$ dependence. This is also true for Wilson loops with dependence on $\hat{n}_{1,i}$ and not on $X_1$.

Case 2: By making use of BCFW recursion for a single Wilson loop the term $\langle\mathcal{L}[X_1,1,\ldots,\hat{n}_{1,\tilde{\imath}}]\rangle$ appearing in $\tilde{T}_i$ can be expressed as
\begin{align}
    &\langle \mathcal{L}[X_1,1,\hdots,n_1-1,\hat{n}_{1,\tilde{\imath}}]\rangle = \langle \mathcal{L}[1,\hdots,n_1-1,\hat{n}_{1,\tilde{\imath}}]\rangle\notag \\
    &+\sum_{j=2}^{n_1-2} [n_1-1,\hat{n}_{1,\tilde{\imath}},X_1,j-1,j]\langle \mathcal{L}[X_1,1,\hdots,j-1,I_j']\rangle\langle\mathcal{L}[I_j',j,\hdots,n_1-1,\widehat{(\hat{n}_{1,\tilde{\imath}})}_j]\rangle\,,
\label{1wlbcfw}
\end{align}
where $I_j'=(j-1 \hspace{1mm} j)\cap (n_1-1 \hspace{1mm} \hat{n}_{1,\tilde{\imath}} \hspace{1mm} X_1)$ and $\widehat{(\hat{n}_{1,\tilde{\imath}})}_j=(n_1-1 \hspace{1mm} \hat{n}_{1,\tilde{\imath}}) \cap (j-1 \hspace{1mm} j \hspace{1mm} X_1)$ and each Wilson loop is tree-level. The first term is of a form already discussed and its twistors can be sent to their collinear limit. The $R$-invariant factor in the second term is $O(\epsilon)$ and as this is multiplied by another $O(\epsilon)$ $R$-invariant factor in (\ref{BCFW}), $\chi_{X_1}$ integration will not give rise to an $O(\epsilon^{-2})$ pole so this term vanishes.

Case 3: Also appearing in $\tilde{T}_i$ we find $\langle\mathcal{L}[X_1,1,\ldots,\hat{n}_{1,\tilde{\imath}},I_{\tilde{\imath}},\tilde{\imath},\ldots,\tilde{\imath}-1,I_{\tilde{\imath}}]\rangle^{(k,0)}$ which has diagrams where $X_1$ and $\hat{n}_{1,\tilde{\imath}}$ do not appear together in $R$-invariants and ones where they do. The former are $O(\epsilon^0)$ and if any of the $\chi_{X_1}$ integration acts on it, $\epsilon$ poles are not produced, so the twistors can be sent to their collinear limit. By making use of Pl\"{u}cker relations, it can be seen that the latter diagrams vanish when multiplied by $[n_1-1,n_1,X_1,\tilde{\imath}-1,\tilde{\imath}]$ due to their Grassmann structure\footnote{These diagrams which vanish upon multiplication by $[n_1-1,n_1,X_1,\tilde{\imath}-1,\tilde{\imath}]$ are divergent diagrams of the form described in the previous footnote. This divergence arises from the intersecting limit, not the collinear limit.}. Therefore, the twistors in the Wilson loop can be sent to their collinear limit.

Thus, only the $R$-invariant prefactors $[n_1-1,n_1,X_1,i-1,i]$ and $[n_1-1,n_1,X_1,\tilde{\imath}-1,\tilde{\imath}]$ are subjected to the collinear integration and the correlators can be sent to their collinear limit. This allows us to perform the collinear integration for arbitrary MHV degree.

We have the following collinear limits of the twistors appearing in the BCFW expression,
\begin{alignat}{2}
        & I_2|_{\epsilon\rightarrow 0} = Z_1 \equiv\bar{I}_2\nonumber\\
        & I_i|_{\epsilon\rightarrow 0}= (i-1 \hspace{1mm} i) \cap (n_1-1 \hspace{1mm}n_1 \hspace{1mm}1)\equiv\bar{I}_i, \hspace{5mm} &&i=3,\ldots, n_1-2\nonumber\\
        &I_{\tilde{\imath}}|_{\epsilon\rightarrow 0}=(\tilde{\imath}-1 \tilde{\imath})\cap (n_1-1 \hspace{1mm} n_1 \hspace{1mm} 1) \equiv \bar{I}_{\tilde{\imath}},\hspace{5mm}&&\tilde{\imath}=\tilde{1},\ldots,\tilde{n}_2\nonumber\\
        & Z_{\hat{n}_{1,i}}|_{\epsilon\rightarrow 0}=Z_{n_1}, \hspace{5mm} &&i=1,\ldots,n_1 \nonumber\\
        &Z_{\hat{n}_{1,\tilde{\imath}}}|_{\epsilon\rightarrow 0}=Z_{n_1}, \hspace{5mm}&&\tilde{\imath}=\tilde{1},\ldots,\tilde{n}_2 \nonumber\\
        & Z_X|_{\epsilon\rightarrow 0}=Z_{n_1}
        \label{twistorlimits}
\end{alignat}

We now look at the action of the collinear integration on the $R$-invariant factors. If an $R$-invariant is not of the form $[X_1,i,j,n_1-1,n_1]$, $[X_1,1,i,j,n_1]$, $[X_1,1,i,n_1-1,n_1]$, $[i,j,k,n_1,X_1]$, where $i,j,k\neq n_1-1,1$, then it integrates to 0 \cite{Caron-Huot:2011dec},
\begin{equation}
\int \bigl[d^{2|3}Z_{X_1}\bigr]^{A'}_A [X_1,a,b,c,d]=0\,.
\label{vanishingRinv}
\end{equation}
Using the BCFW expression for the correlator of two Wilson loops given in (\ref{BCFW}), the expression for the $X_1$-dependent single Wilson loop (\ref{contribterm}), the limits of the twistors (\ref{twistorlimits}) appearing in the correlators of the BCFW expression and the fact that $R$-invariants not of the form given above integrate to 0, we can write the right-hand side of (\ref{qbar1loop}) as
\begin{align}
    \biggl[&\sum_{i=3}^{n_1-2} \int \bigl[d^{2|3}Z_{X_1}\bigr]_A^{A'}[i-1,i,n_1-1,n_1,X_1]\notag \\
    &\times\Big[\sum_{k'=2}^k\bigl(\langle\mathcal{L}[n_1,1,\ldots,i-1,\bar{I}_i]\mathcal{L}[\tilde{1},\ldots,\tilde{n}_2]\rangle^{(k',0),{\rm c}}\langle\mathcal{L}[\bar{I}_i,i,\ldots,n_1-1,n_1]\rangle^{(k-k',0)} \notag \\
    &\hspace{13mm}+\langle\mathcal{L}[n_1,1,\ldots,i-1,\bar{I}_i]\rangle^{(k'-k,0)}\langle\mathcal{L}[\tilde{1},\ldots,\tilde{n}_2] \mathcal{L}[\bar{I}_i,i,\ldots,n_1-1,n_1]\rangle^{(k',0),{\rm c}}\bigr) \notag \\
    &\hspace{13mm}-\langle\mathcal{L}[1,\ldots,n_1]\mathcal{L}[\tilde{1},\ldots,\tilde{n}_2]\rangle^{(k,0),{\rm c}}\Big] \notag\\
    +&\sum_{\tilde{\imath}=\tilde{1}}^{\tilde{n}_2} \int \bigl[d^{2|3}Z_{X_1}\bigr]_A^{A'}[\tilde{\imath}-1,\tilde{\imath},n_1-1,n_1,X_1]\notag \\
    &\times\Big[\langle\mathcal{L}[n_1,1,\ldots,n_1,\bar{I}_{\tilde{\imath}},\tilde{\imath},\ldots,\tilde{\imath}-1,\bar{I}_{\tilde{\imath}}]\rangle^{(k,0)}-\sum_{k'=0}^{k}\langle\mathcal{L}[1,\ldots,n_1]\rangle^{(k',0)}\langle\mathcal{L}[\tilde{1},\ldots,\tilde{n}_2]\rangle^{(k-k',0)}\Big] \notag \\
    &\hspace{10mm}
+\text{cyc}_1 \biggr] + \text{loop}_1 \leftrightarrow \text{loop}_2\,.
\label{qbarsimpleRinv}
\end{align}
Here, we also used
\begin{equation}
\begin{split}
    &\langle\mathcal{L}[1,\ldots,n]\rangle^{(0,0)}=1\\
    &\langle\mathcal{L}[1,\ldots,n]\mathcal{L}[\tilde{1},\ldots,\tilde{m}]\rangle^{(0,0),{\rm c}}=0\\
    &\langle\mathcal{L}[1,\ldots,n]\mathcal{L}[\tilde{1},\ldots,\tilde{m}]\rangle^{(1,0),{\rm c}}=0
\end{split}
\end{equation}
In (\ref{qbarsimpleRinv}), the first term of the BCFW expansion of the 2 Wilson loop correlator is combined with the $X_1$-dependent single Wilson loop term. The $j=2$ term of each of these cancel. Crucially, as the entirety of the collinear integration is now isolated to single $R$-invariants, it is trivial to perform the collinear integration regardless of the MHV degree being inspected.

The non-zero integrals of the $R$-invariants mentioned above (\ref{vanishingRinv}) are given in \cite{Caron-Huot:2011dec}. The only integral we require is
\begin{equation}
    \int \bigl[d^{2|3}Z_{X_1}\bigr]^{A'}_A [i,j,n_1-1,n_1,X_1]=\int^{\tau=\infty}_{\tau=0}d\hspace{0.5mm}\textrm{log}\frac{\langle Y \hspace{1mm}i \hspace{1mm}j\rangle}{\langle Y\hspace{1mm}n_1-2 \hspace{1mm} n_1-1\rangle}\bar{Q}^{A'}_A \hspace{0.5mm} \textrm{log}\frac{\langle\bar{n}_1 \hspace{1mm}j\rangle}{\langle \bar{n}_1\hspace{1mm} i \rangle}
    \label{Rinvint}
\end{equation}
where we have the bitwistor $Y=n\wedge B$, $Z_B=Z_{n_1-1}-C\tau Z_1$, $(\bar{n}_1)=(n_1-1 \hspace{1mm} n_1 \hspace{1mm} 1)$ and $i,j \neq 1, n_1-1$. Substituting this integral in (\ref{qbarsimpleRinv}), we see that there is a $\tau=\infty$ pole in each term due to the presence of $d\hspace{0.5mm}\text{log}\langle Y \hspace{1mm} n_1-2 \hspace{1mm} n_1-1 \rangle$. This pole non-trivially cancels out in the sum. To see that this pole does not appear in the full expression, a BCFW expression of the first term of the RHS of the $\bar{Q}$-equation can be obtained by instead using the shift $Z_{X_1} \rightarrow Z_{X_1}+t Z_1$. This expression has no terms with $d\textrm{log}\langle Y \hspace{1mm} n_1-2 \hspace{1mm} n_1-1 \rangle$, so there are no $\tau=\infty$ poles. However, it will have a $d\hspace{0.5mm}\textrm{log}\langle Y \hspace{1mm} 1 \hspace{1mm} 2 \rangle$ in each term, which has a pole at $\tau=0$. Due to the equality of the BCFW expressions, there are no $\tau$ poles. Thus, we can disregard the $d\hspace{0.5mm}\textrm{log} \langle Y \hspace{1mm} n_1-2 \hspace{1mm} n_1-1 \rangle$ term in (\ref{Rinvint}),
\begin{equation}
    \int^{\tau=\infty}_{\tau=0} d\hspace{0.5mm}\textrm{log} \langle Y \hspace{1mm}i \hspace{1mm} j\rangle \hspace{1mm}\bar{Q}\hspace{0.5mm} \textrm{log} \frac{\langle\overline{n}_1 \hspace{1mm} j \rangle}{\langle \overline{n}_1 \hspace{1mm} i\rangle}=\textrm{log} \langle Y \hspace{1mm} i \hspace{1mm} j \rangle \hspace{1mm} \bar{Q}\hspace{0.5mm} \textrm{log} \frac{\langle \overline{n}_1 \hspace{1mm} j \rangle}{\langle \overline{n}_1 \hspace{1mm} i \rangle}\Bigg|_{\tau=0}^{\tau=\infty}=\textrm{log} \frac{\langle n_1 \hspace{1mm}1\hspace{1mm}i \hspace{1mm} j\rangle}{\langle n_1-1 \hspace{1mm} n_1 \hspace{1mm} i \hspace{1mm} j \rangle} \bar{Q}\hspace{0.5mm} \textrm{log} \frac{\langle \overline{n}_1 \hspace{1mm} j\rangle}{\langle\overline{n}_1 \hspace{1mm} i\rangle}
\end{equation}

The integrated expression for the RHS of the $\bar{Q}$-equation is thus 
\begin{align}
    \bar{Q}_A^{A'}\mathcal{R}_{n_1,n_2}^{(k,1),c} = \biggl[&\sum_{i=3}^{n_1-2}\textrm{log}\frac{\langle n_1 \hspace{1mm} 1 \hspace{1mm}i-1 \hspace{1mm} i \rangle}{\langle n_1-1 \hspace{1mm}n_1 \hspace{1mm}i-1 \hspace{1mm}i\rangle} \bar{Q}_A^{A'} \textrm{log} \frac{\langle \bar{n}_1 \hspace{1mm}i\rangle}{\langle \overline{n}_1 \hspace{1mm}i-1\rangle}\notag \\
    &\times\Big[\sum_{k'=2}^k(\langle\mathcal{L}[n_1,1,\ldots,i-1,\bar{I}_i]\mathcal{L}[\tilde{1},\ldots,\tilde{n}_2]\rangle^{(k',0),{\rm c}}\langle\mathcal{L}[\bar{I}_i,i,\ldots,n_1-1,n_1]\rangle^{(k-k',0)}\notag \\
    &+\langle\mathcal{L}[n_1,1,\ldots,i-1,\bar{I}_i]\rangle^{(k'-k,0)}\langle \mathcal{L}[\tilde{1},\ldots,\tilde{n}_2] \mathcal{L}[\bar{I}_i,i,\ldots,n_1-1,n_1]\rangle^{(k',0),c})\notag \\
    &-\langle\mathcal{L}[1,\ldots,n_1]\mathcal{L}[\tilde{1},\ldots,\tilde{n}_2]\rangle^{(k,0),{\rm c}}\Big]\notag \\
    +&\sum_{\tilde{\imath}=\tilde{1}}^{\tilde{n}_2}\textrm{log}\frac{\langle n_1 \hspace{1mm} 1 \hspace{1mm}\tilde{\imath}-1 \hspace{1mm} \tilde{\imath} \rangle}{\langle n_1-1 \hspace{1mm}n_1 \hspace{1mm}\tilde{\imath}-1 \hspace{1mm}\tilde{\imath}\rangle} \bar{Q}_A^{A'} \textrm{log} \frac{\langle \bar{n}_1 \hspace{1mm}\tilde{\imath}\rangle}{\langle \overline{n}_1 \hspace{1mm}\tilde{\imath}-1\rangle}\notag \\
    &\times\Big[\langle\mathcal{L}[n_1,1,\ldots,n_1,\bar{I}_{\tilde{\imath}},\tilde{\imath},\ldots,\tilde{\imath}-1,\bar{I}_{\tilde{\imath}}]\rangle^{(k,0)} \notag \\
    &-\sum_{k'=0}^{k}\langle\mathcal{L}[1,\ldots,n_1]\rangle^{(k',0)}\langle\mathcal{L}[\tilde{1},\ldots,\tilde{n}_2]\rangle^{(k-k',0)}\Big] + \text{cyc}_1\biggr]+ \text{loop}_1 \leftrightarrow \text{loop}_2
\label{qbarintegrated}
\end{align}

At this stage both the left-hand side and right-hand side of (\ref{qbar1loop}) are given as a sum over terms of the form \(\log(f)\bar{Q}_A^{A'}\log(g)\) dressed by coefficients which can be probed at the appropriate MHV degree. We have verified numerically that the two sides are equal for all cases where the one-loop correlators on the left-hand side are at N${}^2$MHV or N$^3$MHV, with all combinations of multiplicities up to and including eight sides on each Wilson loop.

\section{Conclusion}
The correlation functions of multiple light-like Wilson loops are kinematically complicated objects. However, we have shown here that the $O(g^2)$ problem is entirely tractable through the use of the chiral box expansion, and indeed in \cite{leadingSing} will present the fully-general $O(g^2)$ solution for any number of Wilson loops of any multiplicity, and at all MHV degree. This data has provided convincing checks on our generalised \(\bar{Q}\)-equation which we may now confidently assert holds in the $SU(N)$ theory. 

An obvious and pressing question is the feasibility of $O(g^4)$ and even $O(g^6)$ calculations. At e.g. $O(g^4)$, although there are bases of local integrals in the literature \cite{Bourjaily:2015jna}, in general the integrated expressions are not known for these. However, the \(\bar{Q}\)-equation has been used to great success in computing single Wilson loops up to \(O(g^6)\) \cite{Caron-Huot:2011dec,Li:2021bwg,He:2020vob,Caron-Huot:2013vda}. For instance, by computing the action of the collinear integral on the \(O(g^2)\), N\({}^2\)MHV contribution to a correlator of a pentagon and a square, it should be possible to compute the action of \(\bar{Q}\) on the \(O(g^4)\), NMHV contribution to a square-square correlator, and thus to obtain this integrated correlator. Starting from an \(O(g^2)\) N\(^2\)MHV pentagon-pentagon and hexagon-square, it should likewise be possible to obtain the MHV square-square correlator at \(O(g^6)\). It will be interesting to determine at which stage elliptic integrals might arise in this procedure.

Investigating the $\bar{Q}$-equation as above would provide interesting data which may provide clues regarding the analytic structure of Wilson loop correlators. In the case of a single Wilson loop expectation value (or equivalently a planar scattering amplitude), the symbol alphabets are related to cluster algebras \cite{Golden:2013xva}, with further constraints on the analytic structure via \emph{cluster adjacency} \cite{Drummond:2017ssj}. While the modified cyclic symmetry for these objects will clearly modify the ordinary notion of the positive region of the Grassmannian which applies for a single Wilson loop, it would be interesting to investigate to what extent cluster-like structures are applicable in this more general setting. 

The \(\bar{Q}\)-equation which we have verified here may also be useful for the study of the correlator of a Wilson loop with a local (chiral) Lagrangian, much studied in the literature. It was observed in \cite{Chicherin:2022zxo} that at two-loops, the symbol of the five point result exhibits stringent restrictions on the final entries, which are evocative of a \(\bar{Q}\)-equation for these objects. It would be very interesting if such a \(\bar{Q}\)-equation be extracted by taking an appropriate limit of the \(\bar{Q}\)-equation which we have exhibited here for the correlators of multiple Wilson loops.

\section*{Acknowledgements}
The authors thank \"Omer G\"urdo\u gan for discussions and collaboration at the early stages of this project. All authors are supported by the STFC consolidated grant ST/X000583/1.

\appendix

\newpage

\section{Explicit formulae for chiral box integrands and integrals}
\label{chiralBoxes}
Here for convenience we collect explicit formulae for the integrands and integrated expressions which feature in the chiral box basis of one-loop local integrals. We choose our conventions such that the first integrand has a cut of \(1\) on the corresponding Schubert solution, and the second integrand has a cut of \(-1\) on the relevant Schubert solution, where in the residue computation one puts the poles in the order they are listed in when labelling the chiral box. In practice the Mathematica package supplied in the ancilliary files of \cite{Bourjaily:2013mma} is a highly effective means of quickly generating these integrands and integrals.  

\textbf{Zero mass}, \((ij)(jk)(kl)(li)\)

Schubert solution 1: \((AB) = (jl)\)

Integrand 1: 
\[\frac{1}{\pi^2}\frac{\langle ijkl \rangle \langle X jl \rangle \langle ABik \rangle}{\langle ABij \rangle \langle ABjk \rangle \langle ABkl \rangle \langle ABli \rangle \langle ABX \rangle }\]

Schubert solution 2: \((AB) = (ik)\)

Integrand 2:
\[\frac{1}{\pi^2}\frac{\langle ijkl \rangle \langle X ik \rangle \langle ABjl \rangle}{\langle ABij \rangle \langle ABjk \rangle \langle ABkl \rangle \langle ABli \rangle \langle ABX \rangle }\]

Integral:
\[
-3\Li_2(1) - \frac{1}{2}\log\bigl(\frac{\langle ijX \rangle \langle klX \rangle}{\langle jkX \rangle \langle ilX \rangle}\bigr)^2\]

\textbf{One mass} \((ij)(jk)(kl)(lm)\)

Schubert solution 1: \((AB) = (jl)\)

Integrand 1: 
\[\frac{1}{\pi^2}\frac{\langle AB (ijk)\cap(klm) \rangle \langle Xjl \rangle }{\langle ABij \rangle \langle ABjk \rangle \langle ABkl \rangle \langle ABlm \rangle \langle ABX \rangle}\]

Schubert solution 2: \((AB) = (ijk) \cap (klm)\)

Integrand 2: 
\[\frac{1}{\pi^2}\frac{\langle X (ijk)\cap(klm) \rangle \langle ABjl \rangle }{\langle ABij \rangle \langle ABjk \rangle \langle ABkl \rangle \langle ABlm \rangle \langle ABX \rangle}\]

Integral: 
\[-\Li_{2}(1) + \Li_2\bigl(1 - \frac{\langle ijX \rangle \langle jklm \rangle}{\langle jk X \rangle \langle ijlm \rangle}\bigr) + \Li_2\bigl(1 - \frac{\langle ijkl \rangle \langle lm X \rangle }{\langle kl X \rangle \langle ijlm \rangle}\bigr) + \log\bigl(\frac{\langle ijX \rangle \langle jklm \rangle}{\langle jkX \rangle \langle ijlm \rangle}\bigr)\log\bigl(\frac{\langle ijkl \rangle \langle lm X \rangle}{\langle kl X \rangle \langle ijlm \rangle}\bigr)\]

\textbf{Two mass easy} \((ij)(jk)(lm)(mn)\)

Schubert solution 1: \((AB) = (jm)\)

Integrand 1: 
\[\frac{1}{\pi^2}\frac{\langle AB (ijk) \cap (lmn) \rangle \langle X jm \rangle}{\langle AB ij \rangle \langle ABjk \rangle \langle ABlm \rangle \langle ABmn \rangle  \langle ABX \rangle }\]

Schubert solution 2: \((AB) = (ijk) \cap (lmn)\)

Integrand 2:  
\[\frac{1}{\pi^2}\frac{\langle X  (ijk) \cap (lmn) \rangle \langle AB jm \rangle}{\langle AB ij \rangle \langle ABjk \rangle \langle ABlm \rangle \langle ABmn \rangle \langle ABX \rangle }\]

Integral: \begin{align}
-&\Li_2\bigl(1-\frac{\langle ij X \rangle \langle jklm \rangle}{\langle jk X \rangle \langle ijlm \rangle}\bigr) + \Li_2\bigl(1 - \frac{\langle ijX \rangle \langle jkmn \rangle }{\langle jkX\rangle \langle ijmn \rangle}\bigr) + \Li_2\bigl(1 - \frac{\langle jklm \rangle \langle ijmn \rangle}{\langle jkmn \rangle \langle ijlm \rangle}\bigr) \notag \\
-&\Li_2\bigl(1-\frac{\langle jklm \rangle \langle mn X \rangle}{\langle jkmn \rangle \langle lm X \rangle}\bigr) + \Li_2\bigl(1-\frac{\langle ijlm \rangle \langle mnX \rangle }{\langle lmX \rangle \langle ijmn \rangle}\bigr) + \log\bigl(\frac{\langle ijX\rangle \langle jkmn\rangle}{\langle jkX\rangle \langle ijmn \rangle}\bigr)\log\bigl(\frac{\langle ijlm \rangle \langle mnX\rangle}{\langle lmX\rangle \langle ijmn \rangle}\bigr) \notag
\end{align}

\textbf{Two mass hard} \((ij)(jk)(kl)(mn)\)

Schubert solution 1: \((AB) = (ijk) \cap (kmn)\)

Integrand 1: 
\[
\frac{1}{\pi^2}\frac{\langle X (jkl) \cap ( mn (ij) \cap (ABk)) \rangle - \langle X (jmn) \cap ( kl (ij) \cap (ABk))\rangle}{2\langle ABij\rangle \langle ABjk \rangle \langle ABkl \rangle \langle ABmn \rangle \langle ABX \rangle}
\]

Schubert solution 2: \((AB) = (mnj) \cap (jkl)\)

Integrand 2:
\[\frac{1}{\pi^2}\frac{\langle X (kmn) \cap (ij (kl) \cap (ABj)) \rangle - \langle X (ijk) \cap (mn (kl) \cap (ABj)) }{2\langle ABij \rangle \langle ABjk \rangle \langle ABkl \rangle \langle ABmn \rangle \langle ABX \rangle }\]

Integral:
\begin{align} 
- \Li_2\bigl(1 - \frac{\langle jkX \rangle \langle klmn \rangle}{\langle jkmn \rangle \langle klX \rangle } \bigr) &+ \Li_2\bigl(1- \frac{\langle ijX \rangle\langle jkmn \rangle}{\langle jkX \rangle \langle ijmn \rangle}\bigr) - \frac{1}{2}\log\bigl(\frac{\langle ijX\rangle \langle jkmn \rangle}{\langle jkX\rangle \langle ijmn\rangle}\bigr)\log\bigl(\frac{\langle klX\rangle \langle ijmn \rangle}{\langle ijkl \rangle \langle mnX\rangle}\bigr)\notag \\
& -\frac{1}{2}\log\bigl(\frac{\langle jkX \rangle \langle klmn \rangle}{\langle jkmn \rangle \langle klX \rangle}\bigr)\log\bigl(\frac{\langle ijkl \rangle \langle mnX \rangle}{\langle klX\rangle \langle ijmn \rangle}\bigr)\notag
\end{align}

\textbf{Three mass} \((ij)(jk)(lm)(no)\)

Schubert solution 1: \((AB) = (jlm) \cap (noj)\)

Integrand 1: 
\[\frac{1}{\pi^2}\frac{\langle X (jlm) \cap (no (ijk) \cap (AB) \rangle - \langle X (jno) \cap (lm (ijk) \cap (AB)) \rangle}{2 \langle ABij \rangle \langle ABjk \rangle \langle ABlm \rangle \langle ABno \rangle \langle ABX \rangle}\]

Schubert solution 2: \((A,B) = ((ijk) \cap (lm), (no) \cap (ijk))\)

Integrand 2:
\[\frac{1}{\pi^2}\frac{\langle X (lm) \cap (ijk) (no) \cap (ABj) \rangle - \langle X (no) \cap (ijk) (lm) \cap (ABj) \rangle }{2 \langle ABij \rangle \langle ABjk \rangle \langle ABlm \rangle \langle ABno \rangle \langle ABX \rangle}\]

Integral:
\begin{align}
-&\Li_2\bigl(1-\frac{\langle ijX \rangle \langle jklm \rangle }{\langle jkX \rangle \langle ijlm \rangle}\bigr) - \Li_2\bigl(1-\frac{\langle jkX\rangle \langle ijno \rangle}{\langle ijX \rangle \langle jkno\rangle }\bigr) + \Li_2\bigl(1 - \frac{\langle jklm \rangle \langle ijno \rangle}{\langle jkno \rangle \langle ijlm \rangle}\bigr) \notag \\
+&\frac{1}{2}\log\bigl(\frac{\langle ijX \rangle \langle jklm \rangle}{\langle jkX \rangle \langle ijlm \rangle }\bigr)\log\bigl(\frac{\langle jkX \rangle \langle lmno \rangle}{\langle jkno \rangle \langle lm X \rangle}\bigr)  + \frac{1}{2}\log\bigl(\frac{\langle jkX \rangle \langle ijno \rangle}{\langle ijX \rangle \langle jkno \rangle}\bigr) \log\bigl(\frac{\langle ijX\rangle \langle lmno \rangle}{\langle ijlm \rangle \langle no X \rangle}\bigr)\notag
 \end{align}

\textbf{Four mass} \((ij)(kl)(mn)(op)\)

Schubert solution 1: \(+\) choice for \(\Delta\)

Integrand 1: 
\[\frac{1}{\pi^2}\frac{\frac{1}{2} \langle ijmn \rangle \langle klop \rangle \langle ABX \rangle \Delta - \frac{1}{12}\epsilon^{\alpha \beta \gamma \delta} \langle AB L_{\alpha} \cap (L_{\beta}X_1) L_{\gamma} \cap (L_{\delta}X_2)  \rangle}{\langle ABij \rangle \langle ABkl \rangle \langle ABmn \rangle \langle ABop \rangle \langle ABX \rangle}\]

where we define
\[(L_1,L_2,L_3,L_4) = ((ij),(kl),(mn),(op)).\]

Schubert solution 2: \(-\) choice for \(\Delta\)

Integrand 2:
\[\frac{1}{\pi^2}\frac{\frac{1}{2} \langle ijmn \rangle \langle klop \rangle \langle ABX \rangle \Delta + \frac{1}{12}\epsilon^{\alpha \beta \gamma \delta} \langle AB L_{\alpha} \cap (L_{\beta}X_1) L_{\gamma} \cap (L_{\delta}X_2)  \rangle}{\langle ABij \rangle \langle ABkl \rangle \langle ABmn \rangle \langle ABop \rangle \langle ABX \rangle}\]

Integral:
\begin{align}
&\Li_2\bigl(\frac{1}{2}(1 - u + v + \Delta )\bigr)  +\Li_2\bigl(\frac{1}{2}(1 + u - v + \Delta )\bigr) - \Li_2(1) \notag \\
&+ \frac{1}{2}\log(u)\log(v) - \log\bigl(\frac{1}{2}(1 - u + v + \Delta )\bigr)\log\bigl(\frac{1}{2}(1 + u - v + \Delta )\bigr)   \notag 
\end{align}

where we define
\[\Delta = \sqrt{(1-u-v)^2-4uv}\]
\[u= \frac{\langle ijkl \rangle \langle mnop \rangle}{\langle ijmn \rangle \langle klop \rangle}\]
and
\[v= \frac{\langle klmn \rangle \langle ijop\rangle}{\langle ijmn \rangle \langle klop \rangle}.\]

\section{Integrating the RHS of colour-exact $\bar{Q}$ equation}
\label{rhsqbarnonplanar}

This computation closely follows the procedure of \cite{Caron-Huot:2011dec} and Section \ref{rhsqbarplanar} but instead makes use of a colour-exact Wilson loop recursion relation which we will derive and discuss in more detail in forthcoming work \cite{multiBCFW}; the derivation closely follows \cite{Bullimore:2011ni}.\\

The $O(g^2)$ part of the $\bar{Q}$ equation for arbitrary $k$ is given by
\be
\bar{Q}_A^{A'} \mathcal{R}^{(k,1)}_{n} = \frac{2C_F}{N} \int [d^{2|3} \mathcal{Z}_{X}]^{A'}_A \bigl[ \mathcal{R}^{(k+1,0)}_{n+1} - \mathcal{R}^{(k,0)}_n \mathcal{R}^{(1,0)}_{n+1}\bigr] + \text{ cyc}\,.
\label{Qbarnpanyk}
\ee
The LHS can be evaluated using the chiral box expansion of the remainder function. $\bar{Q}$ annihilates the box coefficients and thus only acts on the chiral boxes, which are made up of logs and dilogs, upon which $\bar{Q}$ acts in a simple way. More details on this are given in Section \ref{lhsqbar}.\\

To make the RHS of (\ref{Qbarnpanyk}) more explicit, let us include the labels of the vertices
\begin{equation}
    \begin{split}
      &\bar{Q}_A^{A'} \mathcal{R}^{(k,1)}_{n} \\
      &=\frac{2C_F}{N} \int [d^{2|3} \mathcal{Z}_{X}]^{A'}_A \bigl[ \langle \mathcal{L}[1,\hdots,n,X]\rangle^{(k+1,0)} - \langle \mathcal{L}[1,\hdots,n]\rangle^{(k,0)}\langle \mathcal{L}[1,\hdots,n,X]\rangle^{(1,0)}\bigr] + \text{ cyc}\,.
\label{Qbarnpanyk2}  
    \end{split}
\end{equation}
$Z_{X}$ is parameterised as follows,
\begin{equation}
    Z_{X}=Z_{n}-\epsilon Z_{n-1}+ C \epsilon \tau Z_1 +C' \epsilon^2 Z_2
\end{equation} 
where $C=\frac{\langle n-1 \hspace{1mm} n \hspace{1mm} 2 \hspace{1mm} 3 \rangle}{\langle n \hspace{1mm} 1 \hspace{1mm} 2 \hspace{1mm} 3\rangle}$ and $C'=\frac{\langle n-2 \hspace{1mm} n-1 \hspace{1mm} n \hspace{1mm} 1 \rangle}{\langle n-2 \hspace{1mm} n-1 \hspace{1mm} 2 \hspace{1mm} 1 \rangle}$.

There are two objects that need to be integrated, $\langle \mathcal{L}[1,\hdots,n_1,X]\rangle^{(k+1,0)}$ and $\langle \mathcal{L}[1,\hdots,n_1,X]\rangle^{(1,0)}$. The integration passes through objects without $X$ dependence. From a diagrammatic expansion, we have 
\begin{equation}
        \langle\mathcal{L}[1,\ldots,n_1,X]\rangle^{(1,0)}=\sum_{i=1}^{n_1-1}\sum_{\substack{j=i+2,\\j \neq i-1,i,i+1}}^{X}[*,i-1,i,j-1,j]
        \label{singleWL2}
\end{equation}
Choosing $*=n-1$ and noting that $n$ and $X$ must appear together to contribute to the collinear integral, the only contribution to this factor is
\begin{equation}
    \sum_{i=2}^{n-2}[n-1,i-1,i,n,X].
\end{equation}
Collinear integration acts on these $R$-invariants to give expressions of the form $\log(f)\bar{Q}^{A'}_A \log(g)$, as detailed in \cite{Caron-Huot:2011dec}. To integrate the other object on the right-hand side, we use a recursion relation to isolate the $\epsilon$ poles into a single $R$-invariant so that the collinear integration only acts on this factor. The colour-exact recursion relation for a single Wilson loop which we make use of is given by
\begin{equation}
\begin{split}
    &\langle \mathcal{L}[1,\hdots,n,X]\rangle^{(k+1,0)} = \langle \mathcal{L}[1,\hdots,n]\rangle^{(k+1,0)}\\
    &+\sum_{i=2}^{n-2} [n-1,n,X,i-1,i]\bigl(\langle \mathcal{L}[X,1,\hdots,i-1,I_i]\mathcal{L}[I_i,i,\hdots,n-1,\hat{n}_i]\rangle^{(k,0)}\\
    &\hspace{37.25mm}-\frac{\alpha}{N^2}\langle \mathcal{L}[X,1,\hdots,i-1,I_i,i,\hdots,n-1,\hat{n}_i,I_i]\rangle^{(k,0)}\bigr)
\end{split}
\label{nonplanarbcfw}
\end{equation}
where
\begin{equation}
    I_i=(i-1\hspace{1mm} i)\cap(n-1\hspace{1mm}n\hspace{1mm}X), \hspace{5mm} Z_{\hat{n}_i}=(n-1\hspace{1mm} n)\cap (i-1\hspace{1mm}i\hspace{1mm}X)
\end{equation}
and each correlator is colour exact. In this relation, $\alpha=0$ for $U(N)$ and $\alpha=1$ for $SU(N)$ which is the case of present interest. Using the same logic as explained below \ref{intersectpoints}, the collinear integration only acts on the R-invariant factor and the twistors in the correlators are simply sent to their collinear limit. We have the following collinear limits,
    \begin{alignat}{2}
    &I_2|_{\epsilon\rightarrow 0}=Z_1 \equiv \bar{I}_2 &&\\
    &I_i|_{\epsilon\rightarrow 0}=(i-1 \hspace{1mm}i)\cap (n-1 \hspace{1mm} n \hspace{1mm}1)\equiv \bar{I}_i, \hspace{7mm} &&i=3,\hdots,n-2\\
    &Z_{\hat{n}_i}|_{\epsilon\rightarrow 0}=Z_{n},\hspace{7mm} &&i=1,\hdots,n
    \end{alignat}
After applying the collinear limit to the Wilson loops, it can be seen that the $i=2$ term in \ref{nonplanarbcfw} cancels against the $i=2$ term in (\ref{singleWL2}) when that is plugged into the second term on the RHS of (\ref{Qbarnpanyk2}). Using the integrated expression for the $R$-invariant (\ref{Rinvint}), noting the cancellation of poles in the same manner described below (\ref{Rinvint}) and the cancellation of the $i=2$ terms in the sum of the first and second terms in (\ref{Qbarnpanyk}), we obtain the integrated expression for the RHS of the $\bar{Q}$ equation,
\begin{equation}
\begin{split}
    &\bar{Q}_A^{A'} \mathcal{R}_n^{(k,1)}\\
    &=\frac{2C_F}{N}\sum_{i=3}^{n-2} \log\frac{\langle n \hspace{1mm} 1 \hspace{1mm} i-1 \hspace{1mm} i \rangle}{\langle n-1 \hspace{1mm} n \hspace{1mm} i-1 \hspace{1mm} i \rangle}\bar{Q}\log \frac{\langle n-1 \hspace{1mm} n \hspace{1mm} 1 \hspace{1mm} i\rangle}{\langle n-1 \hspace{1mm} n \hspace{1mm} 1 \hspace{1mm} i-1\rangle }\\
    &\hspace{4mm}\times\bigl(\langle \mathcal{L}[n,1,\hdots,i-1,\bar{I}_i]\mathcal{L}[\bar{I}_i,i,\hdots,n]\rangle^{(k,0)}-\frac{\alpha}{N^2} \langle \mathcal{L}[n,1,\hdots,i-1,\bar{I}_i,i,\hdots,n,\bar{I}_i]\rangle^{(k,0)}\\
    &\hspace{10mm}+ \left(1-\frac{1}{N^2}\right)\langle \mathcal{L}[1,\hdots,n]\rangle^{(k,0)}\bigr)+\textrm{cyc}
\end{split}
\end{equation}
The $O(g^2)$ Wilson loop on the LHS can be computed by taking cuts of the integrand obtained from a diagrammatic expansion and multiplying these by the corresponding chiral boxes, as outlined in section \ref{compleadingsings}. $\bar{Q}$ annihilates the leading singularities and only acts on the boxes, as in (\ref{qbaract1}) and (\ref{qbaract2}). Then, we may straightforwardly verify numerically that the left and right hand side agree. 

\appendix


\end{document}